\documentclass[aps,prl,onecolumn,showpacs,superscriptaddress,groupedaddress,footinbib]{revtex4-1}

\usepackage{amsmath}
\usepackage{bm}
\usepackage[dvipdfmx]{graphicx}
\usepackage{mathrsfs}
\usepackage{color}
\usepackage[normalem]{ulem}
\DeclareMathOperator\erfc{erfc}

\begin{document}


\title{Chromatin remodeling due to transient-link-and-pass activity enhances subnuclear dynamics}
\author
{\parbox{\linewidth}{\centering
Rakesh Das$^{1,\ast, \#}$, Takahiro Sakaue$^{2}$, G. V. Shivashankar$^{3,4}$, Jacques Prost$^{1,5,\dagger}$ and \\ Tetsuya Hiraiwa$^{1,6,\star}$\\
\normalsize{$^{1}$ Mechanobiology Institute, National University of Singapore, Singapore 117411}\\
\normalsize{$^{2}$ Department of Physical Sciences, Aoyama Gakuin University, Kanagawa 252-5258, Japan}\\
\normalsize{$^{3}$ Department of Health Sciences and Technology (D-HEST), ETH Zurich, Villigen 8092, Switzerland}\\
\normalsize{$^{4}$ Division of Biology and Chemistry, Paul Scherrer Institute, Villigen 5232, Switzerland}\\
\normalsize{$^{5}$ Laboratoire Physico Chimie Curie, Institut Curie, Paris Science et Lettres Research University, 75005 Paris, France}\\
\normalsize{$^{6}$ Institute of Physics, Academia Sinica, Taipei City 115201, Taiwan}\\
\normalsize{$^{\ast}$ rakeshd68@yahoo.com};
\normalsize{$^{\dagger}$ Jacques.Prost@curie.fr};
\normalsize{$^{\star}$ thiraiwa@gate.sinica.edu.tw}\\
\normalsize{$^{\#}$ Current address: Max Planck Institute for the Physics of Complex Systems, N\"othnitzer Strasse 38, 01187 Dresden, Germany}.
\vspace{5mm}
}}

\date{\today}


\begin{abstract}
%
Spatiotemporal coordination of chromatin and subnuclear compartments is crucial for cells. 
Numerous enzymes act inside 
nucleus\textemdash some of those transiently link and pass two chromatin segments.
Here we 
study how such
an active perturbation
affects fluctuating dynamics of an inclusion in the chromatic medium. 
%
Using numerical simulations and a versatile effective model, we
categorize inclusion dynamics into three distinct modes. The 
transient-link-and-pass
activity speeds up 
inclusion dynamics by affecting a slow mode 
related to
chromatin remodeling, viz., size and shape of the chromatin meshes. 
\end{abstract}

\maketitle


Genetic information of a eukaryotic cell is stored in its chromatin, a $\sim2$ m long polymeric entity comprising DNA and histone proteins, which is packed inside the nucleus typically of size $7$-$10$ $\mu$m.
In addition to the chromatin, the nucleus contains a diverse variety of subnuclear compartments (SNCs) like nucleoli, speckles, Cajal bodies, promyelocytic leukemia bodies, transcription factories etc., ranging $\sim50$ nm - $1$ $\mu$m in sizes, all dispersed in a viscous fluid medium called nucleoplasm \cite{ZidovskaBiophysRev2020, ShivaBook2010, KosakBiochimBiophysActa2014, SpectorTrendsGenet2011, BrangwynneCurrOpinCellBiol2015}. 
A spatiotemporal coordination among these SNCs and the chromatin is necessary for the healthy functionality of the cell \cite{ZidovskaBiophysRev2020, ShivaBook2010, KosakBiochimBiophysActa2014, ShabanGenomeBiol2020, MisteliCurrOpinGenetDev, BrangwynneCurrOpinCellBiol2015, SuterTrendsCellBiol2020} lack of which correlates with several diseases \cite{SabariDevCell2020, SleemanCurrOpinCellBiol2015}.  

Recent studies have attributed such spatiotemporal coordination of SNCs and chromatin to the mechanical state of chromatin.  
Ref.~\cite{BrangwynneNatPhys2021} shows that coalescence kinetics of inert liquid droplets dispersed inside a nucleus depends on their dynamics dictated by the mechanics of the chromatic environment, whereas Ref.~\cite{WingreenPRL2021} shows how the number, size, and localization of such subnuclear condensates are dictated by chromatin mechanics. 
We found in our earlier study \cite{RakeshELife2022} that, 
transient-link-and-pass activity (TLPA)
associated with ATP-dependent actions of some classes of enzymes, like Topoisomerase-II \cite{KouzineTranscription2013, HsiehAnnuRevBiochem2013, RocaNuclAcidRes2009, NitissNatRevCancer2009}, can affect the microphase-separation structure of hetero-/eu-chromatin\textemdash this enables us to speculate that, even in a homogeneous medium of chromatin, {\it e.g.}, even when looking into only the euchromatin part, enzymatic activities could affect local mechanical states of chromatin.
Such change in the local mechanical state of chromatin could eventually affect dynamics of finite-size inclusions such as SNCs. 
Indeed, it is known that the dynamics of the chromatin and 
other SNCs are usually ATP-dependent \cite{ShivaBiophysJ2008, ShivaPlosOne2012, SwedlowNatCellBiol2002, GorischPNAS2004, ZidovskaPNAS2013, ZidovskaPNAS2018, DiPierroPNAS2018, ShiNatComm2018}, and ATP-dependent remodeling of the chromatic environment is recognized as one of the mechanisms for this dependency \cite{ShivaBiophysJ2012, GorischPNAS2004, ShivaPlosOne2012}.
Despite its possible relevance for biological 
functions, up to present, no studies have clarified if 
an active perturbation like TLPA
actually plays a role in 
inclusion dynamics 
and what kind of chromatic remodeling can contribute to it. 

\begin{figure}[t]
\includegraphics[width=0.6\linewidth]{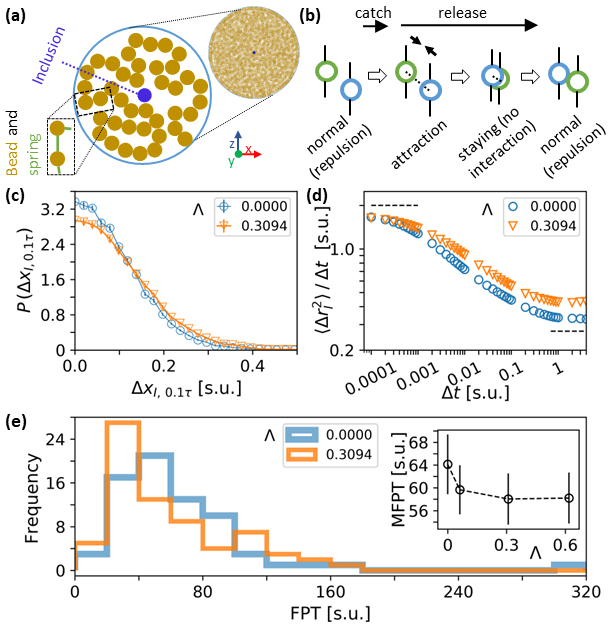}
\caption{ 
(a) Bead-and-spring homopolymer model of chromatin packed inside a spherical cavity together with an inclusion at the center; a schematic and a typical simulation snapshot are shown.
(b) Catch-and-release mechanism of Topoisomerase-II's enzymatic activity. There is no steric repulsion between the pair of the beads when bound to the enzyme. 
(c) Mean $\pm$ s.e.m. of $\Delta x_{I,\,0.1\tau}$-distribution are shown. Only the positive half is shown for better representation.
(d) MSDs of the inclusion are shown for several $\Lambda$. Dashed lines are to guide the early-time and the late-time diffusion. 
(e) Main\textemdash FPT distribution of the inclusion to the radius $R=5\,\ell$ of the cavity starting from $R=0$ ($n=71$).
Inset\textemdash $\Lambda$-dependency of the mean $\pm$ s.e.m. of FPT.
SFIG.~1 supplements FIG.~\ref{Fig1}c\textendash e with curves for four $\Lambda$-values. 
}
\label{Fig1}
\end{figure}

In this Letter, we 
study 
the effects of an active perturbation
on the chromatic medium and fluctuating dynamics of an inclusion, and investigate the underlying physical mechanisms.
For this purpose, we consider a single polymer chain  
subjected to TLPA
as a chromatin model, and a finite-size bead disconnected from the polymer chain as an inclusion, inside a spherical cavity (FIG.~\ref{Fig1}a). 
To implement 
TLPA, 
we follow our earlier work in which a model enzymatic activity was constructed imagining Topoisomerase-II \cite{RakeshELife2022}.
Through numerical simulations of 
our model, we first show that 
TLPA 
indeed affects the inclusion dynamics. 
After that, we investigate what kind of chromatin remodeling is associated with this effect. 
Finally, 
we construct an effective model which is a fluctuating free particle model but keeps the essence to reproduce the TLPA-dependency of the inclusion dynamics observed in our simulation.  
Using the effective model, we identify the three major dynamical modes in the system.

So far, experimentally-observed features of SNC dynamics has been explained quantitatively by a combination of its diffusion within the chromatic-interspace region, plus the translocation of that region due to chromatin diffusion (which we refer as fast and normal diffusive modes, respectively, later in the text) \cite{GorischPNAS2004}. 
However, in general, chromatic interspace itself can remodel, {\it i.e.}, its size and shape can change, and thus affect SNC dynamics \cite{ShivaPlosOne2012}. 
In particular, mechanical actions of enzymes like TLPA may directly drive such chromatin remodeling.
The effective model proposed here indeed demonstrates the relevance of a slow dynamical mode linked to chromatin remodeling for the TLPA-dependent inclusion dynamics.


{\it Model}\textemdash 
We develop a self-avoiding linear homopolymer model for the chromatin confined within a spherical cavity of diameter $D$ (FIG.~\ref{Fig1}a).
The homopolymer comprises $N$ soft-core beads of diameter $d_{B}$ consecutively connected by finitely-extensible-nonlinear-elastic springs.
A hard-core spherical inclusion of diameter $d_I$ is placed at the center of the cavity.
This inclusion experiences steric forces due to the surrounding polymer. 

The dynamics of the positions of the beads (${\bm x}_B$) and the inclusion (${\bm x}_I$) are approximated by Brownian dynamics; see Supplemental Material \cite{SM} and Ref.~\cite{RakeshELife2022} for details. 
Stokes' relation has been assumed to mimic the frictional drag due to implicitly considered nucleoplasm.
Setting the thermal energy and the nucleoplasmic viscosity to unity, and $D=12 \, \ell$, we obtain the simulation units (s.u.) of length ($\ell$), time ($\tau$), and energy ($e$).

TLPA is implemented in the following sequential manner, which we call catch-and-release mechanism. 
In the normal state, any pair of spatially proximal beads experiences steric repulsion due to each other (self-avoidance potential $h_{vex}>0$, see Supplemental Material \cite{SM}). 
The enzyme can catch that pair of beads with a Poisson rate $\lambda_{ra}$ (FIG.~\ref{Fig1}b), and upon that, the beads attract each other ($h_{vex}<0$).
Next, the attraction between those two beads is turned off with another Poisson rate $\lambda_{an}$, and the beads stay there for a while without any steric interaction among themselves ($h_{vex}=0$). 
Eventually, the enzyme unbinds from the beads with a Poisson rate $\lambda_{nr}$, and the beads return to their normal state with steric repulsion between themselves. 
Therefore, in our model, enzymatic activity is realized by the following sequence of Poisson transitions of the steric interaction between a pair of proximal beads: $\text{state}\left(h_{vex}>0\right) \xrightarrow{\lambda_{ra}} \text{state}\left(h_{vex}<0\right) \xrightarrow{\lambda_{an}} \text{state}\left(h_{vex}=0\right) \xrightarrow{\lambda_{nr}} \text{state}\left(h_{vex}>0\right)$.
These steps allow the beads to pass across each other stochastically and thereby perturb the medium \cite{RakeshELife2022}. 

We parameterize TLPA as $\Lambda = \lambda_{ra}\left( 1/\lambda_{an} + 1/\lambda_{nr} \right)$, which can be tuned in experiments by controlling ATP concentration \cite{WangJBiolChem1993}. 
We choose $\lambda_{an}=16.7\,\tau^{-1}$ and $\lambda_{nr}=500\,\tau^{-1}$ and tune $\lambda_{ra}$ to control the activity. 
For a given $\Lambda$, we integrate the equations of motion of ${\bm x}_B$ and ${\bm x}_I$ using Euler discretization method, and thereby we simulate the dynamics of the polymer and the inclusion. 
This choice of the rates sets a typical timescale 
$t_{TLPA} = \left(1/\lambda_{an}+1/\lambda_{nr}\right)\simeq0.062\,\tau$ 
for which an enzyme catches a pair of beads.
We have checked that by this timescale, a bead typically moves over only $\mathcal{O}(d_B)$. 

Each realization of the simulation starts from a steady-state configuration of a self-avoiding polymer packed inside the cavity together with the inclusion at the center of the cavity.
The polymer and the inclusion follow the Brownian dynamics as described above.
We note that the inclusion eventually touches the cavity boundary; so, we simulate our active polymer model (APM) until the inclusion touches the cavity boundary for the first time. 
The results reported below are obtained from $n$  
number of independent realizations.
(See Supplemental Material \cite{SM} for further details.)

{\it Inclusion dynamics}\textemdash
We first investigate how inclusion dynamics is affected by TLPA by looking into its (i) one-dimensional displacement over a given time duration, (ii) mean-square-displacement (MSD), and (iii) first passage time (FPT) in the chromatic medium. 
Here, we consider an inclusion of size comparable to that of the bead size ($d_I=0.40 \, \ell$, $d_B \simeq 0.43 \, \ell$).

First, we compare the distribution $P\left(\Delta x_{I,\,0.1\tau}\right)$ of one-dimensional displacement of the inclusion over a time duration $\Delta t = 0.1\,\tau > t_{TLPA}$ for several $\Lambda$ (FIG.~\ref{Fig1}c). 
We note that $P\left(\Delta x_{I,\,0.1\tau}\right)$ follows a Gaussian distribution and widens with $\Lambda$.
The distributions appeared to be symmetric around zero, although with some fluctuations. 
However, we did not find any drift in the inclusion dynamics (SFIG.~2).

Next, we calculate MSD of the inclusion as
$\langle \Delta r_I^2 \left(\Delta t\right) \rangle = \langle \lvert \bm{x}_I\left(t_0+\Delta t\right) - \bm{x}_I\left(t_0\right) \rvert^2 \rangle$,
where $\langle\cdot\rangle$ represents average over several $t_0$ and realizations. 
We note (A) a short early-time diffusive regime, (B) an intermediate subdiffusive regime, and (C) a late-time diffusive regime. 
Following this regime-(C), we note a significant slowing down of the inclusion dynamics (data not shown). 
We check that by this time, the inclusion reaches close to the cavity boundary, and although the inclusion does not interact with the boundary, its dynamics is indirectly affected by the polymeric density at that locality which differs from the bulk region. 
The effect of TLPA at that locality  has a different physics than that in the bulk region, and therefore, it will be discussed in the future.
In this Letter, we focus on the bulk region (up to radius $R=5\,\ell$ instead of the whole cavity of radius $R=6\,\ell$).
We show the regimes (A)-(C) of the inclusion MSD in FIG.~\ref{Fig1}d. 
Note that the crossover time from the early-time regime to the intermediate regime increases with $\Lambda$.

FPT represents the time that the inclusion takes to reach $R = 5\,\ell$ for the first time starting its dynamics from $R=0$ at time $t=0$. 
We prepare a histogram of the FPTs noted for several realizations of simulation and note that $\Lambda$ affects it (FIG.~\ref{Fig1}e\textendash main).
Starting from a high value of the mean FPT (MFPT) for $\Lambda=0$, it decreases monotonically up to a moderate $\Lambda$ 
(FIG.~\ref{Fig1}e\textendash inset). 
Taking the displacement distributions, MSDs, and the FPT-data together, we conclude that TLPA enhances the inclusion dynamics. 

\begin{figure}[b]
\includegraphics[width=0.6\linewidth]{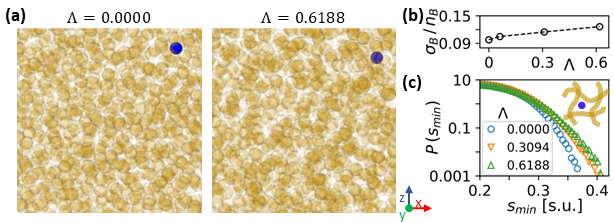}
\caption{ 
(a) Cropped slices ($6\times6$ in s.u.) of typical simulation snapshots are shown for $\Lambda = 0$ (left) and $0.6188$ (right).
(b) Fluctuation $\sigma_B$ in local density $n_B$ of beads increases with $\Lambda$ ($n=71$).
(c) Main\textemdash Distribution of the separation between an arbitrary point and its nearest polymeric-bead. 
$\Lambda$-dependent increase in the largest $s_{min}$ with $P(s_{min}) > 0$ indicates increase in chromatic meshsize. 
Data for $s_{min} < 0.2$ are not shown as the arbitrary points fall on the beads ($d_B\simeq0.43$ s.u.) in that range. 
Inset\textemdash Schematic of the inclusion inside a chromatic mesh.
}
\label{Fig2}
\end{figure}

{\it Chromatic environment}\textemdash
The enhancement of the inclusion dynamics should be attributed to some change in the chromatic environment due to TLPA.
To investigate it in detail, we compare the polymeric configurations with and without TLPA.
We note that the polymeric environment becomes more heterogeneous with $\Lambda$ (FIG.~\ref{Fig2}a, Supplemental~Videos). 
The mean density of the beads, $n_B$, increases with $\Lambda$ (SFIG.~3).
Along with that, the spatiotemporal fluctuation in $n_B$, as manifested by its standard deviation $\sigma_B$, also increases with $\Lambda$ (FIG.~\ref{Fig2}b), suggesting an increase in heterogeneity of chromatin medium with TLPA.

To further elaborate on this observed heterogeneity, we calculate the separation $s_{min}$ of any arbitrary point in the bulk region of the cavity to its nearest bead and prepare its distribution $P(s_{min})$. 
As a corollary to the increasing heterogeneity in chromatic environment, 
we note that the tail of $P(s_{min})$ becomes heavier with $\Lambda$ (FIG.~\ref{Fig2}c).
The largest $s_{min}(\Lambda)$ with non-zero $P(s_{min})$ minus $d_B/2$ could be interpreted as half of the meshsize in the corresponding chromatic environment. 
It is evident from FIG.~\ref{Fig2}c\textendash main that the typical meshsize increases with $\Lambda$.
Considering the $\langle \Delta r_I^2 \rangle$-data, it is straightforward to understand that the inclusion performs early-time diffusion until when it feels the mechanical hindrance due to its chromatic neighborhood beyond which it shows subdiffusion (FIG.~\ref{Fig2}c\textendash schematic).

{\it Effective model}\textemdash
Next, we develop an integrative understanding of chromatic environment-mediated effect of TLPA on inclusion dynamics.
We hypothesize that inclusion dynamics in our APM could be mimicked by its Brownian dynamics with colored noise defined over a coarse-grained timescale $\delta t$.
Therefore, we conceive a coarse-grained effective model\textemdash 
the inclusion is considered alone, and its dynamics is given by
\begin{eqnarray}
\partial_t{\bm x}_{I} = v_{EM} \, {\bm \zeta}_{EM}, \label{EqBrown_EM}
\end{eqnarray}
where the right-hand-side is an effective noise whose characteristics can be defined from the $\Lambda$-dependent behavior of the displacement vector $\Delta {\bm x}_{I,\,\delta t}$ defined over $\delta t$. 
More specifically, $v_{EM}^2 = \langle \lvert \Delta {\bm x}_{I,\,\delta t} \rvert^2 \rangle / (\delta t)^2$, and ${\bm \zeta}_{EM}$ is a Gaussian noise with 
zero-mean and autocorrelation
$\langle {\bm \zeta}_{EM}(0) \cdot {\bm \zeta}_{EM}(\Delta t) \rangle = C_{\Delta {\bm x}_{I,\,\delta t}}\left(\Delta t\right) = \langle \sum_{t_0} \left[ \Delta {\bm x}_{I,\,\delta t}\left(t_0\right) \cdot \Delta {\bm x}_{I,\,\delta t}\left(t_0+\Delta t\right)\right] /  $  $\sum_{t_0} \lvert \Delta {\bm x}_{I,\,\delta t}\left(t_0\right)\rvert^2 \rangle$ ($\langle\cdot\rangle$ indicating average over several realizations).
Hereafter, we consider $\delta t = 0.003 \,\tau$ over which the inclusion dynamics should be affected by the dynamics of its chromatic neighborhood (FIG.~\ref{Fig1}d; also see SFIG.~4). 

\begin{figure}[t]
\includegraphics[width=0.6\linewidth]{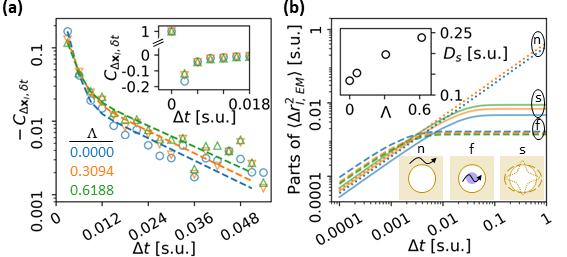}
\caption{ 
(a) Autocorrelation $C_{\Delta {\bm x}_{I,\delta t}}(\Delta t)$ of the inclusion's displacement vector $\Delta {\bm x}_{I,\,\delta t}$ over $\delta t = 0.003\,\tau$ for several $\Lambda$ (symbols). 
Main, semi-log plot to focus on the negative part of $C_{\Delta {\bm x}_{I,\delta t}}$; inset, linear plot.
Dashed-lines are fits to $-C_{\Delta {\bm x}_{I,\delta t}} (\Delta t>0)$ with the double exponential function, $g(\Delta t)$.
(b) Main\textemdash Parts of $\langle \Delta r_{I,\,EM}^2 \rangle$ corresponding to the three dynamical modes (n, f, and s)
for several $\Lambda$ (see FIG.~\ref{Fig3}a for legends). 
Top inset\textemdash 
Diffusivity corresponding to s-mode increases with $\Lambda$. 
Bottom inset\textemdash Schematic of each mode. Golden: chromatin media. Blue: inclusion.
}
\label{Fig3}
\end{figure}

We obtain $C_{\Delta {\bm x}_{I,\,\delta t}}\left(\Delta t\right)$ from our simulation data for several $\Lambda$.
A negative autocorrelation is noted for $\Delta t \ge \delta t$  (FIG.~\ref{Fig3}a\textendash inset) as expected for our current model system\textemdash 
(visco-)elasticity of polymeric medium surrounding the inclusion may tend to reverse the direction of inclusion motion \cite{TakahiroPRE2015}.
The negative part of the autocorrelation shows good fit to double exponential function (FIG.~\ref{Fig3}a\textendash main). 
Therefore, we write 
$C_{\Delta {\bm x}_{I,\delta t}} (\Delta t) = A\delta(\Delta t) - g(\Delta t)$
with $g(\Delta t)=a_{f} e^{- \Delta t / t_{f}} + a_{s} e^{ - \Delta t / t_{s}}$,
where the fitting parameters $a_{f,\, s}$ and $t_{f,\, s}$ depend on $\Lambda$. 
The parameter $A$ takes care of the autocorrelation for $\Delta t < \delta t$. 

Using the effective model, we analytically obtain MSD of the inclusion as
\begin{eqnarray}
\langle \Delta r_{I,\, EM}^2 \left(\Delta t\right) \rangle = \sum_{m \equiv f, s} 2 D_m t_m \left(1 - e^{- \frac{\Delta t}{t_m} }\right)   
                                                              + 2 D_n \Delta t, \;\;\;\;\,\label{EqMSD_EM}
\end{eqnarray}
where $D_{f,s} = v_{EM}^2 a_{f,s} t_{f,s}$ and $D_n = v_{EM}^2 \left( A/2 - a_f t_f - a_s t_s \right)$ (Supplemental Material \cite{SM}).
Treating $A$ as a $\Lambda$-dependent fitting parameter,  we find good agreement between $\langle \Delta r_I^2 \rangle$ and $\langle \Delta r_{I,\, EM}^2 \rangle$ (SFIGs.~5 and 6). 
Thus, our simulation results are successfully reproduced by the effective model, where the chromatic environment-mediated effect of TLPA on the inclusion dynamics is captured by the $\Lambda$-dependency in the features of the coarse-grained noise. 
Below we show that the construction of the effective model allows us to unveil the physical mechanism how TLPA affects the inclusion dynamics.

We identify the MSD-terms corresponding to $m\equiv f, s$ with that obtained for a particle subjected to a harmonic potential following an overdamped Langevin equation (Supplemental Material \cite{SM}). 
This gives us 
a diffusivity $D_m$ and a characteristic time $t_m$ for the mode-$m$.
Thus, the inclusion dynamics can be understood as that dictated by a fast (f) and a slow (s) mode (with $t_f \sim 0.002\;\tau$ and $t_s \sim 0.02\;\tau$), plus a normal (n) diffusive mode.
The n-mode originates from the delta-correlated forces that the inclusion feels due to the thermal noise, plus the polymeric neighborhood coarse-grained over $\delta t$. 
The origin of the fast and slow modes must underlie in the fluctuation in the potential that the inclusion feels due to its polymeric neighborhood;
we indeed found that the fluctuation in the number of the beads interacting with the inclusion has the characteristic timescale comparable to that of the fast mode, whereas remodeling of the polymeric mesh around the inclusion has the timescale comparable with the slow mode (SFIG.~7, Supplemental Material \cite{SM}). 
One can visualize the corresponding scenario as following\textemdash let us consider a polymeric mesh comprising some number of beads around the inclusion. 
At the scale of $\sim 0.002\;\tau$, that number fluctuates because of the individual bead's thermal fluctuation (see SFIG.~8 for bead's MSD), and therefore the potential fluctuates. 
This leads to the fast mode-contribution to inclusion dynamics.
However, at a longer timescale $\sim 0.02\;\tau$, the size and shape of the mesh are reconfigured leading to another source of the potential-fluctuation. 
This dictates the slow mode-contribution to the inclusion dynamics. 


Lastly, we investigate which mode mainly contributes to the speeding of the inclusion dynamics with TLPA. FIG.~\ref{Fig3}b\textendash main plots the parts of MSD, $\langle \Delta r_{I,\, EM}^2 \rangle$, separately derived from all the three modes.
We find that the $\Lambda$-dependency in MSD at the intermediate time scale ($\sim 0.001 <\Delta t < \sim 0.02$) is dominated by that of the s-mode. 
We note a significant increase in $D_s$ with $\Lambda$ (FIG.~\ref{Fig3}b\textendash top inset;  SFIG.~9). 
These results suggest that significant initial speeding up of the inclusion dynamics (around $\Delta t=0.001$ in FIG.~\ref{Fig1}d) is induced through s-mode, which is associated with the chromatin remodeling. In conclusion, TLPA-assisted remodeling of the chromatic neighborhood plays a major role in enhancing the inclusion dynamics.
(See Supplemental Material \cite{SM} for 
discussion on $\Lambda$-dependency of all three diffusivities.)


{\it Discussion}\textemdash
In summary, through numerical simulations of APM and construction and analysis of the effective model, we investigated the effect of enzymatic action-induced TLPA on the inclusion dynamics in the chromatin environment. 
We showed that the inclusion dynamics in this complex medium comprises three modes\textemdash a fast mode dynamics within the local polymeric mesh, a slow mode dynamics associated with the polymeric reconfiguration, and a normal diffusive mode. 
TLPA speeds up the inclusion dynamics by significantly facilitating chromatin remodeling. 
This finding is in line with Ref.~\cite{ShivaPlosOne2012} where the authors emphasized on the importance of ATP-dependent chromatin remodeling to explain their experimental observation. 
In Ref.~\cite{GorischPNAS2004}, as briefly mentioned above, an effective model combining the fast and the normal diffusive modes has been proposed to explain the inclusion dynamics. 
Our results further showed the necessity of the slow mode associated with chromatin remodeling to fully understand 
the inclusion dynamics in active chromatic media
(schematics in FIG.~\ref{Fig3}b\textendash bottom inset).
In Ref.~\cite{ShivaBiophysJ2012}, significant slowing down of transcription compartments' dynamics upon depression of temperature from $37^\circ\text{C}$ to $25^\circ\text{C}$ has been attributed to temperature-dependent active processes. 
The chromatin remodeling-associated dynamical mode reported here can be responsible for that experimental observation. 
Our current study also implicates how a global active perturbation may regulate site-specific target searching of SNCs by enhancing their encounter-frequency \cite{SuterTrendsCellBiol2020, HansenNatChemBiol2020, CortiniNatComm2018}.
Furthermore, as Topoisomerase-II activity is known to correlate with cell cycle \cite{BergerGenes2019}, aging \cite{MullerMutationRes1989}, nucleoplasmic ATP content \cite{YanoSciRep2021}, our study may elucidate such cell state-dependent subnuclear dynamics.

Despite apparent similarity in the inclusion dynamics for various model-scenarios in terms of a coarse-grained quantity like MFPT, we find our effective model useful in distinguishing  the underlying differences in terms of the three dynamical modes (Supplemental Material \cite{SM}) \cite{RD_in_preperation}.

\begin{table}[t]
\begin{center}
\begin{tabular}{|c| c| c| c|} 
  \hline
  {$d_I$} & {$0.2\,\ell$} & {$0.4\,\ell$} & {$1.2\,\ell$} \\
  \hline
  MFPT$_{\Lambda=0.3094}$/MFPT$_{\Lambda=0}$ &$1.04 \pm 0.08$ & $0.90 \pm 0.11$ & $0.57 \pm 0.21$ \\
  \hline
\end{tabular}
\end{center}
\caption{Mean $\pm$ s.e.m. propagated from both the MFPTs.} 
\label{table_MFPT}
\end{table}

Dynamics of an inclusion in a polymeric media is expected to depend on the inclusion size \cite{RubinsteinMacrmolecules2011}.
We find that the effect of TLPA becomes significant for large inclusions (Table~\ref{table_MFPT}, Supplemental Material \cite{SM}) \cite{RD_in_preperation}.

Our model provides a framework which may be helpful for understanding various active gels \cite{LiverpoolPhilTransRSocA2006, JacquesNatPhys2015, MacKintoshScience2007, WeitzCell2014}, usually the long-time stochasticity stemming from activity and the short-time modes being thermal. 
For instance, Refs.~\cite{MacKintoshScience2007, WeitzCell2014} reported time-dependence of inclusion MSDs, which may be related to a reminiscent structural remodeling-associated mode.

\vspace{2mm} 
\begin{acknowledgments}
R.D. acknowledges useful discussions he had with Rapha\"el Voituriez and David Weitz.
R.D. and T.H. appreciate Yuting Lou and Akinori Miyamoto for valuable discussions.
Authors acknowledge anonymous referees for their insightful comments.
This research was supported by Seed fund of Mechanobiology Institute (to J.P., T.H.) and Singapore Ministry of Education Tier 3 grant, MOET32020-0001 (to G.V.S., J.P., T.H.) and JSPS KAKENHI No. JP18H05529 and JP21H05759 from MEXT, Japan (to T.S.). 
\end{acknowledgments}




\clearpage

\onecolumngrid

\renewcommand{\theequation}{S\arabic{equation}}
\renewcommand{\figurename}{SFIG.}
\setcounter{figure}{0}
\setcounter{equation}{0}
\setcounter{section}{0}

\begin{center} 
{\large {\bf Supplemental Material to ``Chromatin remodeling due to transient-link-and-pass activity enhances subnuclear dynamics"}}
\end{center}
\vspace{2mm}

{\centering 
Rakesh Das$^{1,\ast, \#}$, Takahiro Sakaue$^{2}$, G. V. Shivashankar$^{3,4}$, Jacques Prost$^{1,5,\dagger}$ and \\ Tetsuya Hiraiwa$^{1,6,\star}$\\
\normalsize{$^{1}$ Mechanobiology Institute, National University of Singapore, Singapore 117411}\\
\normalsize{$^{2}$ Department of Physical Sciences, Aoyama Gakuin University, Kanagawa 252-5258, Japan}\\
\normalsize{$^{3}$ Department of Health Sciences and Technology (D-HEST), ETH Zurich, Villigen 8092, Switzerland}\\
\normalsize{$^{4}$ Division of Biology and Chemistry, Paul Scherrer Institute, Villigen 5232, Switzerland}\\
\normalsize{$^{5}$ Laboratoire Physico Chimie Curie, Institut Curie, Paris Science et Lettres Research University, 75005 Paris, France}\\
\normalsize{$^{6}$ Institute of Physics, Academia Sinica, Taipei City 115201, Taiwan}\\
\normalsize{$^{\ast}$ rakeshd68@yahoo.com};
\normalsize{$^{\dagger}$ Jacques.Prost@curie.fr};
\normalsize{$^{\star}$ thiraiwa@gate.sinica.edu.tw}\\
\normalsize{$^{\#}$ Current address: Max Planck Institute for the Physics of Complex Systems, N\"othnitzer Strasse 38, 01187 Dresden, Germany}.
}


\tableofcontents


\subsection{Details of the simulation setting} 
We consider a spherical cavity of diameter $D=1.4112$ $\mu$m containing a $50.8070$ Mega base pair (Mbp) long chromatin, yielding a typical diploid genome density of human somatic cells. 
The chromatin is mimicked by a bead-and-spring model comprising $N=20992$ beads, each one representing a $2.4203$ kbp coarse-grained genomic content.  
Assuming nucleosomes as spheres of diameter $22$ nm containing $200$ bp of genomic content and close compaction of nucleosomes within coarse-grained beads, we obtain diameter of the beads as $d_B = 22 \times (\text{number of nucleosomes per bead})^{1/3} = 50.5086$ nm. 

We set $D = 12 \,\ell$ which gives us the simulation unit (s.u.) of length $\ell=117.6$ nm. 
Our model has been simulated at a physiological temperature $T = 310$ K, and we consider energy unit as $e = 1 \,k_BT = 4.28$ pN$\cdot$nm. 
The frictional drag on the beads due to the implicitly considered nucleoplasm is approximated by Stokes' law, and the corresponding viscosity is represented by $\eta$. Considering the nucleoplasmic viscosity as unity in simulations, we get the simulation unit of time $\tau = \eta \ell^3 / e$. 


\subsection{Brownian dynamics} \label{subsec_BD}
We consider the following Brownian dynamics equations to represent the position ${\bm x}_B$ of the $B^{th}$ bead and ${\bm x}_I$ of the inclusion: 
\begin{eqnarray}
\frac{\partial{\bm x}_B}{\partial t} &&= - \frac{1}{6\pi\eta(d_B/2)} \frac{\partial H_B}{\partial {\bm x}} + \sqrt{\frac{2k_BT}{6\pi\eta(d_B/2)}}{\bm \zeta}_B, \label{EqBrown_bead} \\
\frac{\partial{\bm x}_{I}}{\partial t} &&= - \frac{1}{6\pi\eta(d_I/2)} \frac{\partial H_{I}}{\partial {\bm x}} + \sqrt{\frac{2k_BT}{6\pi\eta(d_I/2)}}{\bm \zeta}_{I}. \label{EqBrown_incl}
\end{eqnarray}
Here, ${\bm \zeta}$'s represent univariate white Gaussian noises with zero mean.
We consider the potential energies $H_B$, that realized by the $B^{th}$ bead, and $H_I$, that realized by the inclusion, as 
\begin{eqnarray}
            H_B &=& H_{spring} + H_{vex} + h_{BI} + H_{confinement}  \label{Eq_HB} 
\end{eqnarray}
and
\begin{eqnarray}
            H_I &=& H_{BI}. \label{Eq_HT}
\end{eqnarray}
We explain the terms on the right-hand side of eqs.~(\ref{Eq_HB}) and (\ref{Eq_HT}) sequentially in the following paragraphs.
To numerically integrate eqs.~(\ref{EqBrown_bead}) and (\ref{EqBrown_incl}) with eqs.~(\ref{Eq_HB}) and (\ref{Eq_HT}), time is discretized into steps as usual, and the positions ${\bm x}_B$ of all the beads and ${\bm x}_I$ of the inclusion are updated sequentially over steps. 
 
The potential energy associated with the spring connecting two consecutive beads along the polymer is given by $h_{spring} = - \frac{1}{2} k r_0^2 \ln\left[1-\left(r_{ij}/r_0\right)^2\right]$, where $r_{ij}$ is the distance between the $i^{th}$ and $j^{th}$ beads. 
We consider $H_{spring} = \sum_{j (\in n.i)} h_{spring}$, where the summation $\sum_{j(\in n.i)}$ runs over the beads next to $i$ (i.e., $j$ takes $i-1$ and $i+1$ if the $i^{th}$ bead is not located at one of the polymer ends, whereas it takes only either of them if $i^{th}$ bead is located at an end). Here $k$ is the spring constant of a finitely extensible nonlinear elastic spring whose stretch $r_{ij} \leq r_0$. 

When a pair of beads separated by $r_{ij} \leq \ell$ is {\it not bound} to a Topoisomerase-II (Topo-II) enzyme, each bead realizes steric repulsion due to one another. We consider that interaction potential between those two beads as
\begin{eqnarray}
h_{vex} && =  \epsilon_{vex} \exp\left(-\alpha_{vex} r_{ij}^2\right), \text{  when bead-pair ($ij$) not bound to Topo-II}, 
\label{Hvex1}
\end{eqnarray}
and the total steric potential realized by the $i^{th}$ bead is $H_{vex} = \sum_{j\in r_{ij} \leq \ell} h_{vex}$.
Here $\epsilon_{vex}$ is the amplitude of the Gaussian interaction potential chosen, and $\alpha_{vex}$ determines its variance. 
We set size of the beads in our simulation model by setting a criterion that the minimum of $h_{spring} + h_{vex}$ appears at $r_m = \sum_{i\in \text{connected beads}} d_B / 2$. 
This sets a constraint over the choice of the parameters in $h_{spring}$ and $h_{vex}$ as $k r_m^2 / \left( 1 - r_m^2 / r_0^2 \right) = 2 \epsilon_{vex} \alpha_{vex} r_m^2 \exp\left(-\alpha_{vex}r_m^2\right)$.

As per our active polymer model described in the main text, if a pair of beads ($ij$) separated by $r_{ij} \leq \ell$ is {\it bound} to a Topo-II enzyme, the beads either realize attraction for each-other, or there is no steric interaction among those beads. 
We implement this model scheme by tuning $h_{vex}$ as following:
\begin{eqnarray}
h_{vex}  && = -\epsilon_{vex} \exp\left(-\alpha_{vex} r_{ij}^2\right), \text{  when bead-pair ($ij$) bound to Topo-II and in the attraction state,} \\
         && = 0, \text{when bead-pair ($ij$) bound to Topo-II} \text{  and in the no-interaction state}. 
        \label{Hvex2}
\end{eqnarray}
The interaction potential $h_{vex}$ between a pair of beads ($ij$) separated by $r_{ij} \leq \ell$ starts the following series of Poissonian transitions:
$\text{state}\left(h_{vex}>0\right) \xrightarrow{\lambda_{ra}} \text{state}\left(h_{vex}<0\right) \xrightarrow{\lambda_{an}} \text{state}\left(h_{vex}=0\right) \xrightarrow{\lambda_{nr}} \text{state}\left(h_{vex}>0\right)$,
where each transition step can take place in between any consecutive two steps of ${\bm x}_B$-dynamics (following eq.~(\ref{EqBrown_bead}) and eq.~(\ref{Eq_HB}) discretized in time) and is implemented stochastically according to the Poisson process with the rate $\lambda_{XX}$ ($XX\equiv$ ra,\, an,\, nr). 
When the bead-pair ($ij$) get separated by $r_{ij}\geq \ell$ during this process, those pair of beads temporarily stop proceeding with the stochastic steps of the transitions described above. 

To confine the polymer inside the cavity, we use a potential $H_{confinement}$ that acts upon a bead at a separation $r$ from the boundary of the cavity. We consider this potential energy as that between a wall and a star polymer with functionality $s=2$ (therefore, a linear polymer) \cite{SMJusufiJPCM2001, SMCamargoJCP2009}: 
\begin{eqnarray}
H_{confinement} &=& p s^{3/2} \left[ -\ln\left(\frac{r}{R_s}\right) - \left(\frac{r^2}{R_s^2}-1\right)\left(\frac{1}{1+2\kappa^2R_s^2}-\frac{1}{2}\right) + \gamma \right], \text{  for } r \leq R_s, \notag \\
&=& p s^{3/2} \, \gamma \, {\erfc\left(\kappa r\right)} / {\erfc\left(\kappa R_s\right)} \text{, for } r > R_s.
\label{Hconfinement}
\end{eqnarray}
Here we have modeled individual bead as a star polymer with radius of gyration $R_g = d_B/2$.
In eq.~(\ref{Hconfinement}), $p$ is a free parameter, $R_s = 0.65\,R_g$ stands for radius of corona of the polymer, the parameter $\kappa$ should be  $\mathcal{O}\left(R_g^{-1}\right)$, $\gamma = \sqrt{\pi} \erfc\left(\kappa R_s\right) \exp\left(\kappa^2 R_s^2\right) / \left[\kappa R_s \left(1+2\kappa^2R_s^2\right)\right]$, and $\erfc$ stands for complementary error function.

The steric repulsion between the $B^{th}$ bead and the inclusion, separated by $r_{BI}$, is approximated by WCA potential,
\begin{eqnarray}
h_{BI} &=& 4 \epsilon_{BI} \left[ \left(\frac{\sigma_{BI}}{r_{BI}}\right)^{12} - \left(\frac{\sigma_{BI}}{r_{BI}}\right)^{6} \right] + \epsilon_{BI}, \text{  for } r_{BI} \leq \frac{d_B+d_I}{2}, \notag \\
&=& 0, \text{ otherwise}.
\label{Eq_hBT_WCA}
\end{eqnarray}
Therefore, the total potential energy realized by the inclusion due to neighboring polymeric beads is 
$H_{BI} = \sum_{r_{BI} \leq (d_B+d_I)/2} h_{BI}$.

We summarize our choice of model parameters in Table~\ref{table_param}.

\begin{table}[t]
\begin{center}
\begin{tabular}{|c| c|} 
  \hline
  {\bf Potential} & {\bf Parameters} \\
  \hline
  $H_{spring}$ & $k = 22\, e\ell^{-2}$;  $r_0=2 d_B$. \\
  \hline
  $H_{vex}$ & $\epsilon_{vex}=8\, e$;  $\alpha_{vex}=7.9585\, \ell^{-2}$.\\
  \hline 
  $H_{confinement}$ & $p=4\, e$; $s=2$; $R_s=0.65 R_g$; $\kappa=1/R_g$; $R_g=d_B/2$. \\
  \hline
  $H_{BI}$ & $\epsilon_{BI}=e$; $2^{1/6}\sigma_{BI}=(d_B+d_I)/2$. \\
  \hline
\end{tabular}
\end{center}
    \caption{Choice of model parameters.}
    \label{table_param}
\end{table}

Equations~(\ref{EqBrown_bead}) and (\ref{EqBrown_incl}) are integrated over time in Euler method, where we use a discretized time step $10^{-4} \,\tau$. 
The simulations are done using lab-developed code where we use CUDA to exploit GPU acceleration \cite{SMRakeshELife2022}. 
We start the simulations from specially prepared equilibrated configurations (described in Sec.~\ref{subsec_init_config}) and register the dynamics henceforth.   


\subsection{Preparation of initial configurations} \label{subsec_init_config}
We prepared the initial configurations for our main simulations in two steps.  
We started with a polymer arbitrarily packed inside the spherical cavity, together with the spherical inclusion of diameter $0.8\,\ell$.
We assumed the polymer to be an ideal chain ({\it i.e.}, $h_{vex}=0$ for any pair of the beads), although the interaction between the beads and the inclusion was considered the same as described in Sec.~\ref{subsec_BD}.
We performed Brownian dynamics simulations to evolve the polymeric system for a duration of $4020\,\tau$ (by which mean-squared-displacement of individual bead goes beyond $2D$).
During this simulations, the inclusion was kept fixed at the center of the cavity.
The last configurations obtained from such $12$ set of simulations were considered as the initial configurations for the next step of initial configuration preparation.
In the next step, we considered individual configurations obtained from the first step and replace the previous inclusion by an inclusion particle of desired diameter $d_I$.
This time we considered the polymer to be self-avoiding ({\it i.e.}, $h_{vex}>0$) and allowed the system to evolve for another duration of $4020\,\tau$ (by which the mean-squared-displacement of individual bead goes beyond $D$).
The inclusion was still kept fixed at the center. 
For each set of $12$ initial configurations, we picked $3$ different simulation configurations and consider those $36$ configurations as the initial configurations for our main simulations for a given $d_I$. 
During our main simulations, we have used different sequence of random numbers to generate more than $36$ realizations.

\subsection{Analysis of inclusion's drift dynamics} \label{subsec_drift}
We checked whether there is any emergence of drift motion of the inclusion due to the enzymatic activity. 
Note that we considered an isotropic geometry of the cavity. 
So, by analyzing the distribution of the one-dimensional displacements in one of the three axes of the simulation-reference-frame (SRF), we may not be able to conclude unambiguously the existence of drift. 
So, we did the following. 
For a given realization of simulation, we know the displacement vector ${\bm d}_t = {\bm x}_I(t+0.1\,\tau) - {\bm x}_I(t)$ of the inclusion at time $t$ on the SRF. 
We obtained a transformed vector ${\bm d}_{transformed,\,t} = \mathcal{R}{\bm x}_{I}(t+0.1\,\tau) - \mathcal{R}{\bm x}_{I}(t) \equiv (\Delta x_{t,\,transformed},\Delta y_{t,\,transformed},\Delta z_{t,\,transformed})$ such that $\mathcal{R}{\bm x}_{I}(t) \equiv (0, 0, z_{transformed})$, $\mathcal{R}$ representing necessary sequences of rotations. 
We defined $\Delta z_{transformed} = \langle \Delta z_{t,\,transformed} \rangle_{t \in \text{ course of the simulation}}$ in radial grids (used $6$ concentric radial grids, each of width $\ell$), and similarly $\Delta x_{transformed}$ and $\Delta y_{transformed}$.
We calculated $(\Delta z_{transformed}, \Delta x_{transformed}, \Delta y_{transformed})$'s for $71$ realizations for each $\Lambda$ (small open symbols in SFIG.~\ref{SFig_1D_displacement}), and thereafter we calculated the mean over realizations (big closed symbols in SFIG.~\ref{SFig_1D_displacement}). 
While $\Delta x_{transformed}$ and $\Delta y_{transformed}$ should be zero as per our definitions, a non-zero value of $\Delta z_{transformed}$ would imply a drift. 
However, we do not see any significant non-zero value of $\Delta z_{transformed}$.


\begin{figure}[t]
\includegraphics[width=0.6\linewidth]{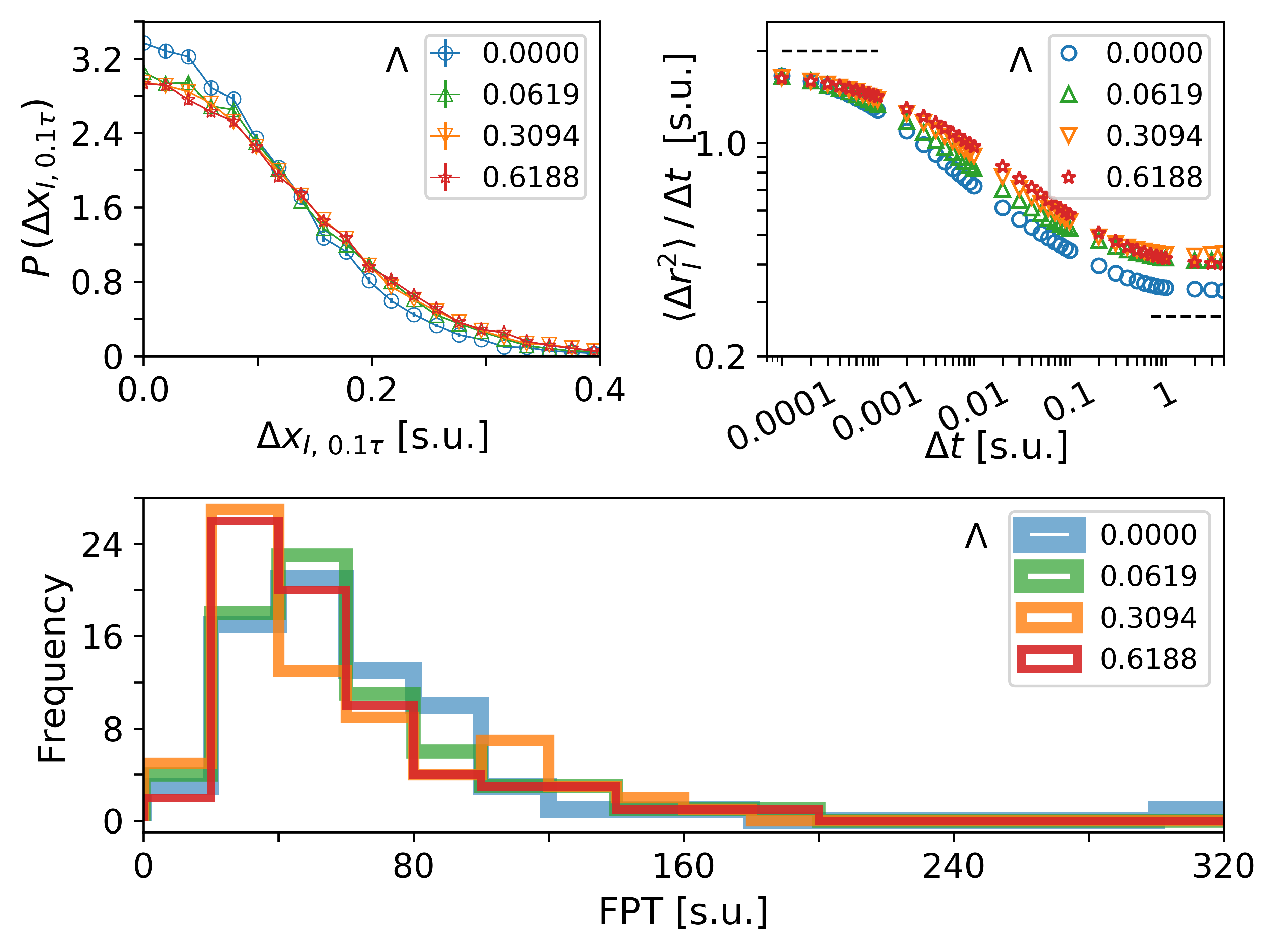}
\caption{ 
Supplementary to Fig.~1(c\textendash e) showing data for four $\Lambda$ values. }
\label{SFigure1cde}
\end{figure}


\begin{figure}[t]
\includegraphics[width=0.6\linewidth]{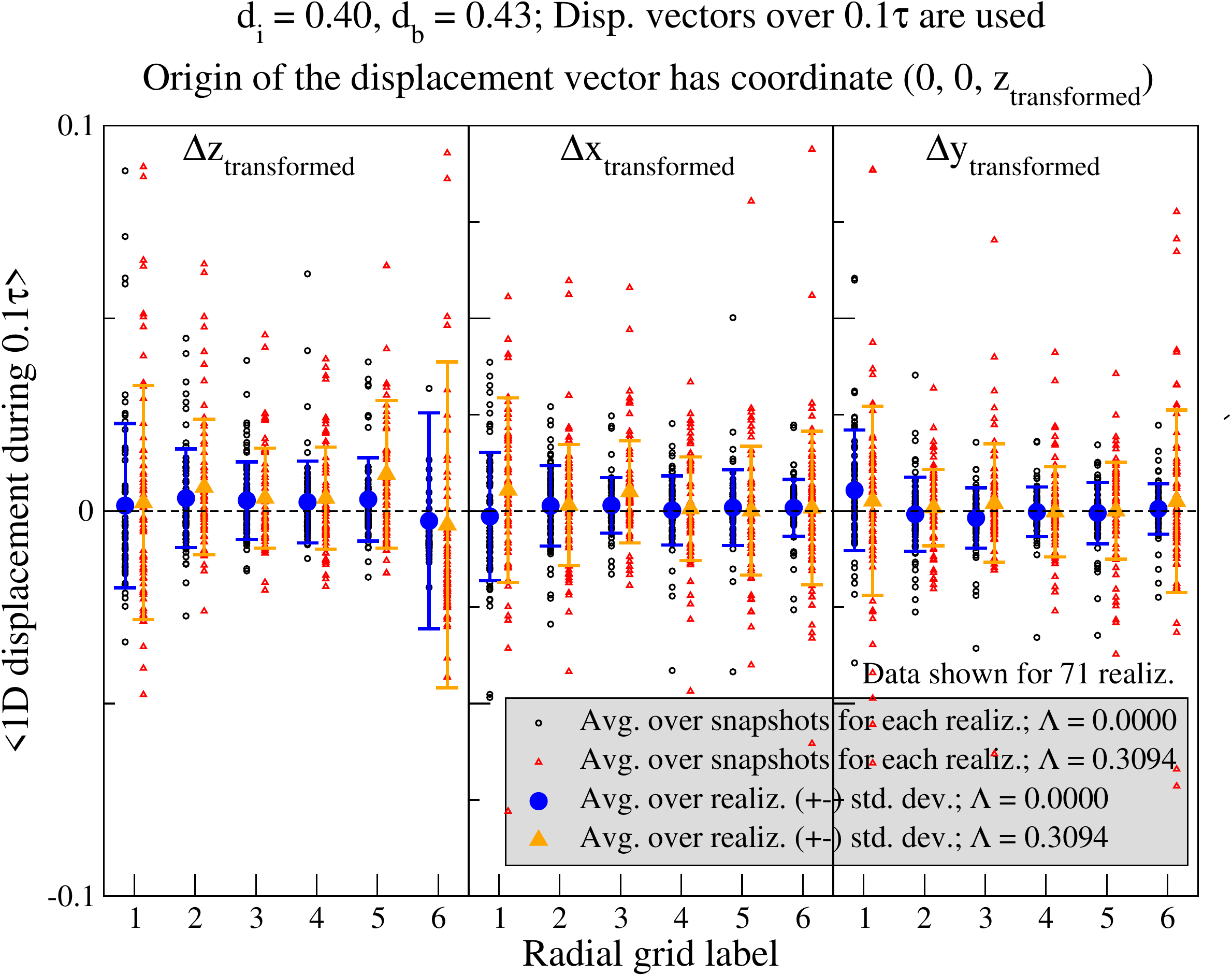}
\caption{ 
We show $(\Delta z_{transformed}, \Delta x_{transformed}, \Delta y_{transformed})$'s for $71$ realizations for each $\Lambda$ by small open symbols. The mean $\pm$ std. dev. over realizations are shown by big closed symbols. We do not see any significant non-zero value of $\Delta z_{transformed}$ inferring no drift motion of the inclusion. }
\label{SFig_1D_displacement}
\end{figure}


\subsection{Density of polymer beads $n_B$} \label{subsec_nB}
\begin{figure}[t]
\includegraphics[width=0.4\linewidth]{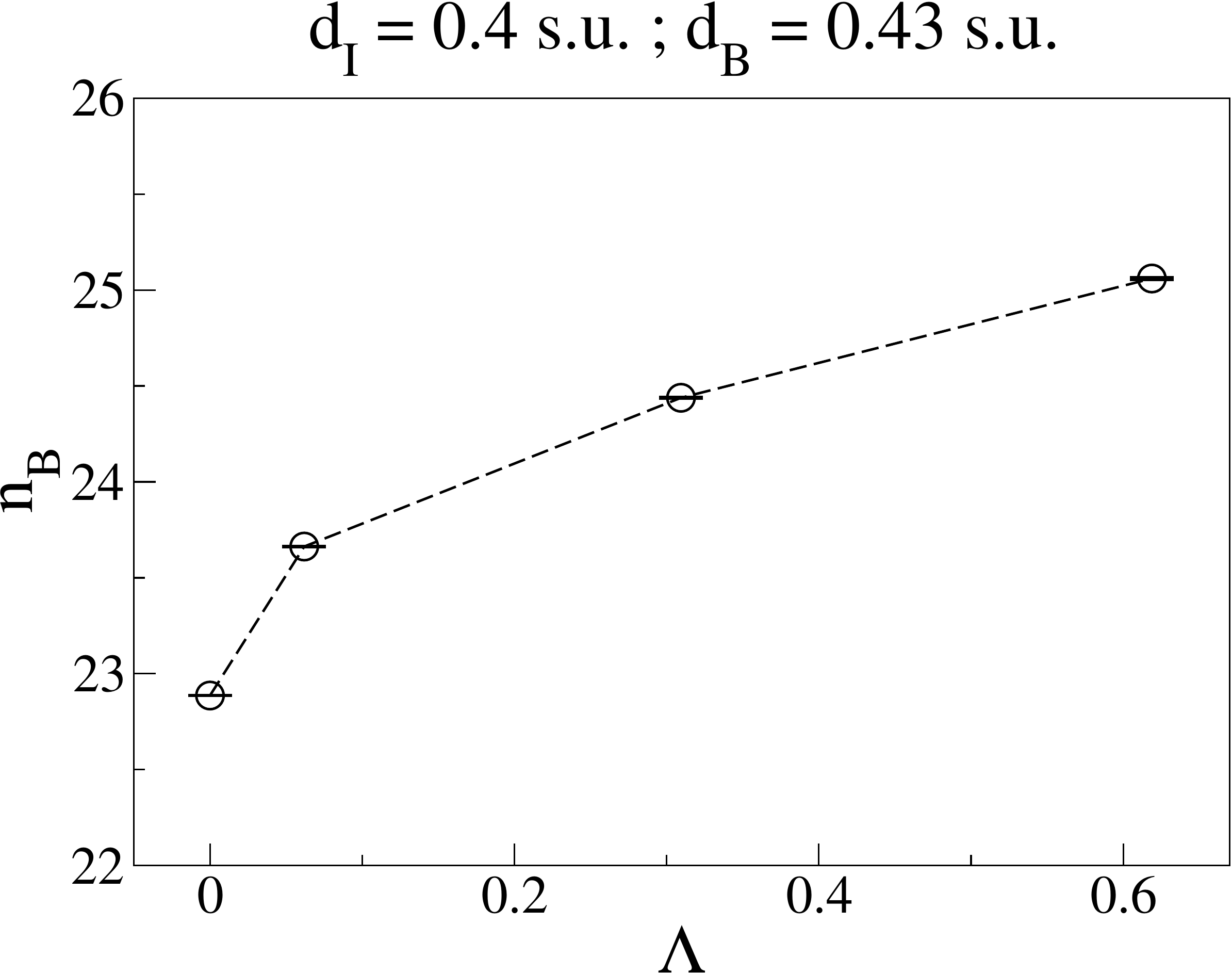}
\caption{ 
Density of polymeric beads, $n_B$, increases with $\Lambda$. Mean $\pm$ s.e.m. over realizations are shown (n=71).}
\label{SFig_poly_bead_density}
\end{figure}

The bulk region of the cavity has been segmented into $5^3$ cubic grids of linear dimension $\ell$, and the number of the polymeric beads in individual grids are counted for several hundreds of snapshots for individual realization. 
First we calculate the mean density of the beads and the corresponding fluctuation for individual realizations, and then taking averages over $71$ realizations, we obtain $n_B$ (SFIG.~\ref{SFig_poly_bead_density}) and $\sigma_B$ (FIG.~2b in the main text).

\subsection{Supplemental Videos}
Two supplemental videos are shown for $\Lambda = 0$ (video~1) and $0.6188$ (video~2). 
In each of the frames, only a slice of breadth $1$ s.u. (around a great circle $y^\prime=0$) of the whole simulation system is shown. 
Each slice (each great circle) is chosen such that it contains the inclusion (blue, $d_I = 0.4$ s.u.) for that frame, and the real simulation coordinates $(x, y, z)$ are transformed accordingly to $(x^\prime, y^\prime, z^\prime)$.
The black square has linear dimension $12$ s.u.

\begin{figure}[b]
\includegraphics[width=0.5\linewidth]{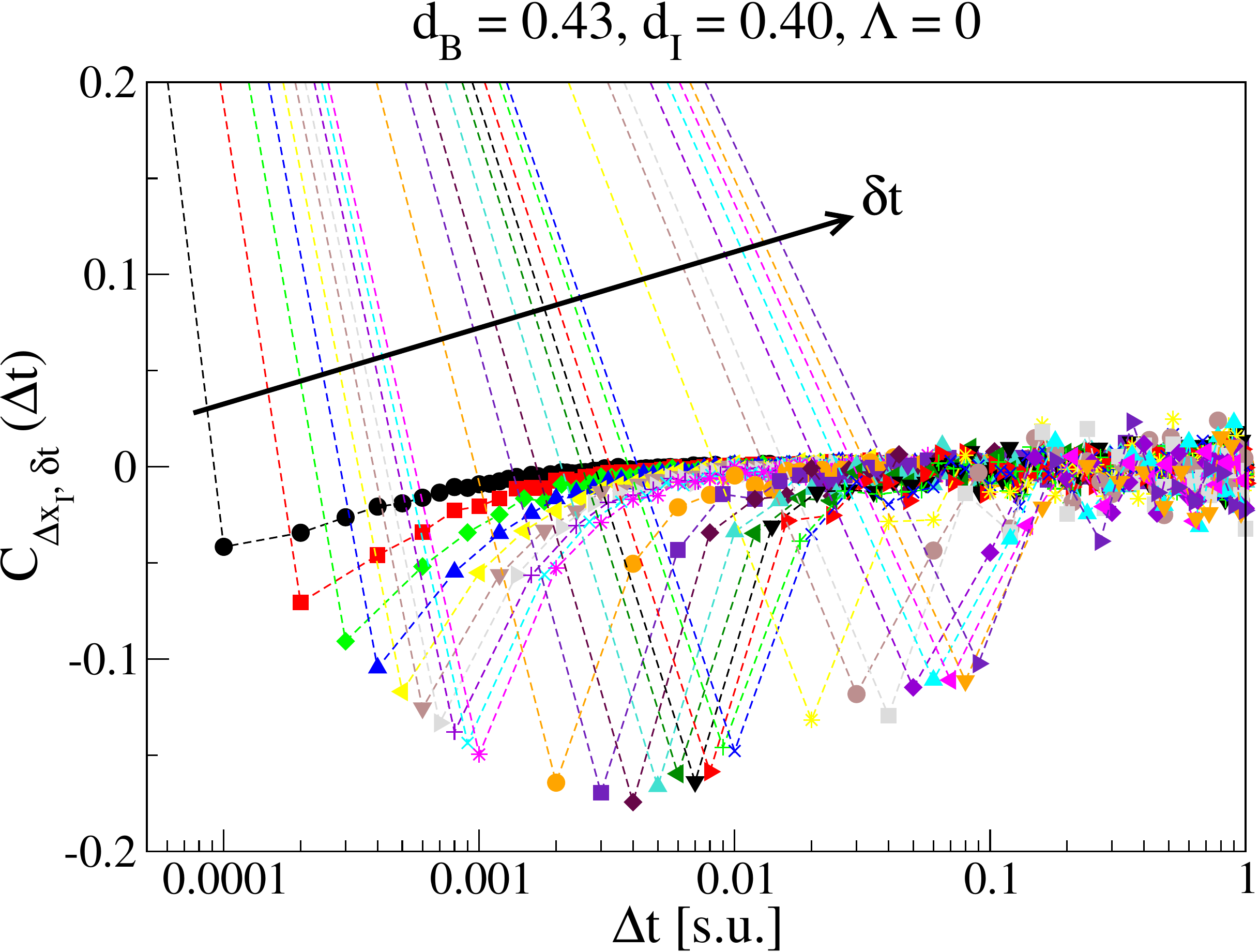}
\caption{ 
Non-monotonic behavior of $C_{\Delta {\bm x}_{I,\,\delta t}}(\Delta t = \delta t)$ with $\delta t$ attaining its optimal about $\mathcal{O}(0.001\,\tau)$.}
\label{SFig_ACF_displacement}
\end{figure}


\subsection{Coarse-grained effective model} \label{subsec_EM}
The autocorrelation function of $\Delta {\bm x}_{I,\,\delta t}$ is defined as 
\begin{equation}
C_{\Delta {\bm x}_{I,\,\delta t}}\left(\Delta t\right) = \frac{\langle \sum_{t_0} \left[ \Delta {\bm x}_{I,\,\delta t}\left(t_0\right) \cdot \Delta {\bm x}_{I,\,\delta t}\left(t_0+\Delta t\right)\right] \rangle}{\langle \sum_{t_0} \lvert \Delta{\bm x}_{I,\,\delta t}\left(t_0\right)\rvert^2 \rangle}. \notag
\end{equation} 
We note a non-monotonic behavior of $C_{\Delta {\bm x}_{I,\,\delta t}}(\Delta t = \delta t)$ with $\delta t$ (SFIG.~\ref{SFig_ACF_displacement}).

\begin{figure}[h]
\includegraphics[width=0.45\linewidth]{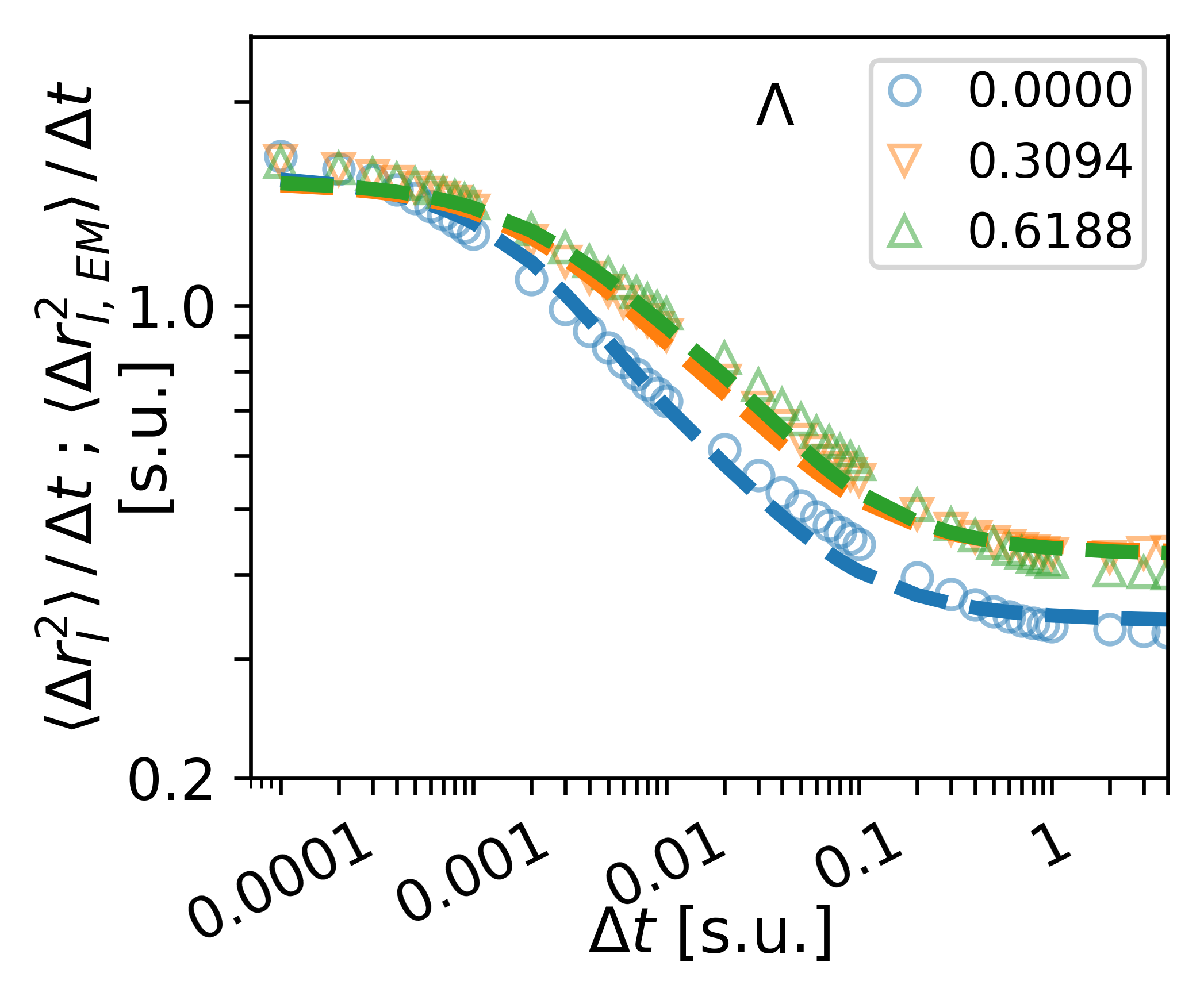}
\caption{The MSD of the inclusion obtained from the simulation model ($\langle \Delta r_I^2 \rangle$, symbols) show excellent match with that obtained using the effective model ($\langle \Delta r_{I,\,EM}^2 \rangle$, dashed-lines).}
\label{SFig_EM_MSDtot}
\end{figure}

\begin{figure}[h]
\includegraphics[width=0.6\linewidth]{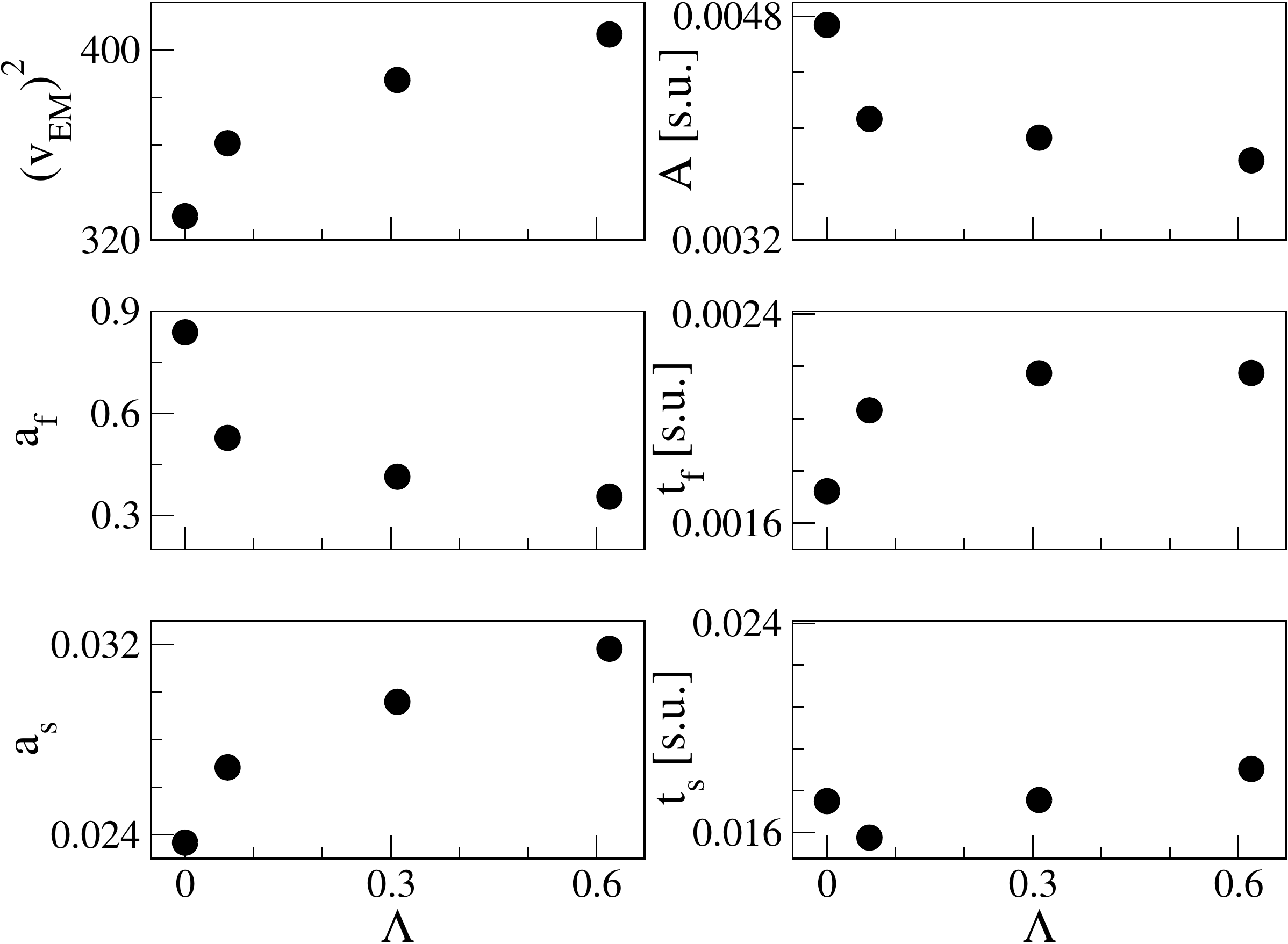}
\caption{Parameters used in the effective model. }
\label{SFig_ACF_fitting}
\end{figure}

The effective model has been defined in the main text by eq.~(1) with the characteristics of the noise $v_{EM}\,{\bm \zeta}_{EM}$ mentioned therein. 
Since $\langle {\bm \zeta}_{EM} \rangle = 0$, the first moment of displacement of the inclusion, $\langle \Delta r_{I,\,EM}(\Delta t) \rangle$ over a time duration $\Delta t$ is zero.
The second moment of the displacement over the duration $\Delta t$ can be written as
\begin{eqnarray}
    \langle \Delta r_{I,\,EM}^2(\Delta t) \rangle &=& v_{EM}^2 \int_t^{t+\Delta t} ds \int_t^{t+\Delta t} ds^\prime \langle {\bm \zeta}_{EM}(s) \cdot {\bm \zeta}_{EM}(s^\prime) \rangle \notag \\
    &=& v_{EM}^2 \int_t^{t+\Delta t} ds \int_t^{t+\Delta t} ds^\prime [\, A\delta(\lvert s-s^\prime \rvert) - a_{f} \exp(- \lvert s-s^\prime \rvert / t_{f}) - a_{s} \exp( - \lvert s-s^\prime \rvert / t_{s}) \,]. 
\end{eqnarray}
A simplification of the above expression leads us to eq.~(2) in the main text. 
Treating $A$ as a $\Lambda$-dependent fitting parameter, we find nice agreement between $\langle \Delta r_{I}^2(\Delta t) \rangle$ and $\langle \Delta r_{I,\,EM}^2(\Delta t) \rangle$ (SFIG.~\ref{SFig_EM_MSDtot}).
The fitted and the used parameters in the effective model are shown in SFIG.~\ref{SFig_ACF_fitting}.

\subsection{Identification of the fast- and slow-modes} \label{subsec_fast_slow_modes_identification}
Let us consider a particle subjected to a harmonic potential performing Brownian motion. 
Its dynamics is given by the following overdamped Langevin equation\textemdash
\begin{eqnarray}
    \frac{\partial {\bm x}_p}{\partial t} = - \frac{1}{\gamma_p} k {\bm x_p} + \sqrt{2 D_p} {\bm \zeta}. \label{Eq_subsec_fast_slow_modes_BD}
\end{eqnarray}
${\bm \zeta}$ represents a Gaussian white noise with $\langle {\bm \zeta} \rangle = 0$ and $\langle {\bm \zeta}(t_1) \cdot {\bm \zeta}(t_2) \rangle = \delta_{\alpha\beta} \delta(t_1-t_2)$; $\alpha, \, \beta \in $ Cartesian components.
The diffusivity is set by the friction coefficient $\gamma_p$ and the thermal energy $k_B T$: $D_p = k_B T / \gamma_p$. 
The spring constant $k$ and friction coefficient $\gamma_p$ set a characteristic timescale $t_c = \gamma_p / k$. 

The solution to eq.~\ref{Eq_subsec_fast_slow_modes_BD} is given by
\begin{eqnarray}
    {\bm x}_p(t) = \sqrt{2 D_p} \int_{-\infty}^{t} ds \; \exp\left(- \frac{t-s}{t_c} \right) \; {\bm \zeta}(s). \label{Eq_subsec_fast_slow_modes_soln}
\end{eqnarray}
It is straightforward to calculate the MSD of the particle\textemdash
\begin{eqnarray}
    \text{MSD}_p(\Delta t) = 2 D_p t_c \left[ 1 - \exp\left( - \Delta t / t_c \right) \right]. \label{Eq_subsec_fast_slow_modes_MSD}
\end{eqnarray}
This expression is analogous to the fast- and the slow-mode contributions to the inclusion MSD, as discussed in the main text; see eq.~(2). 

\begin{figure}[b]
\includegraphics[width=0.43\linewidth]{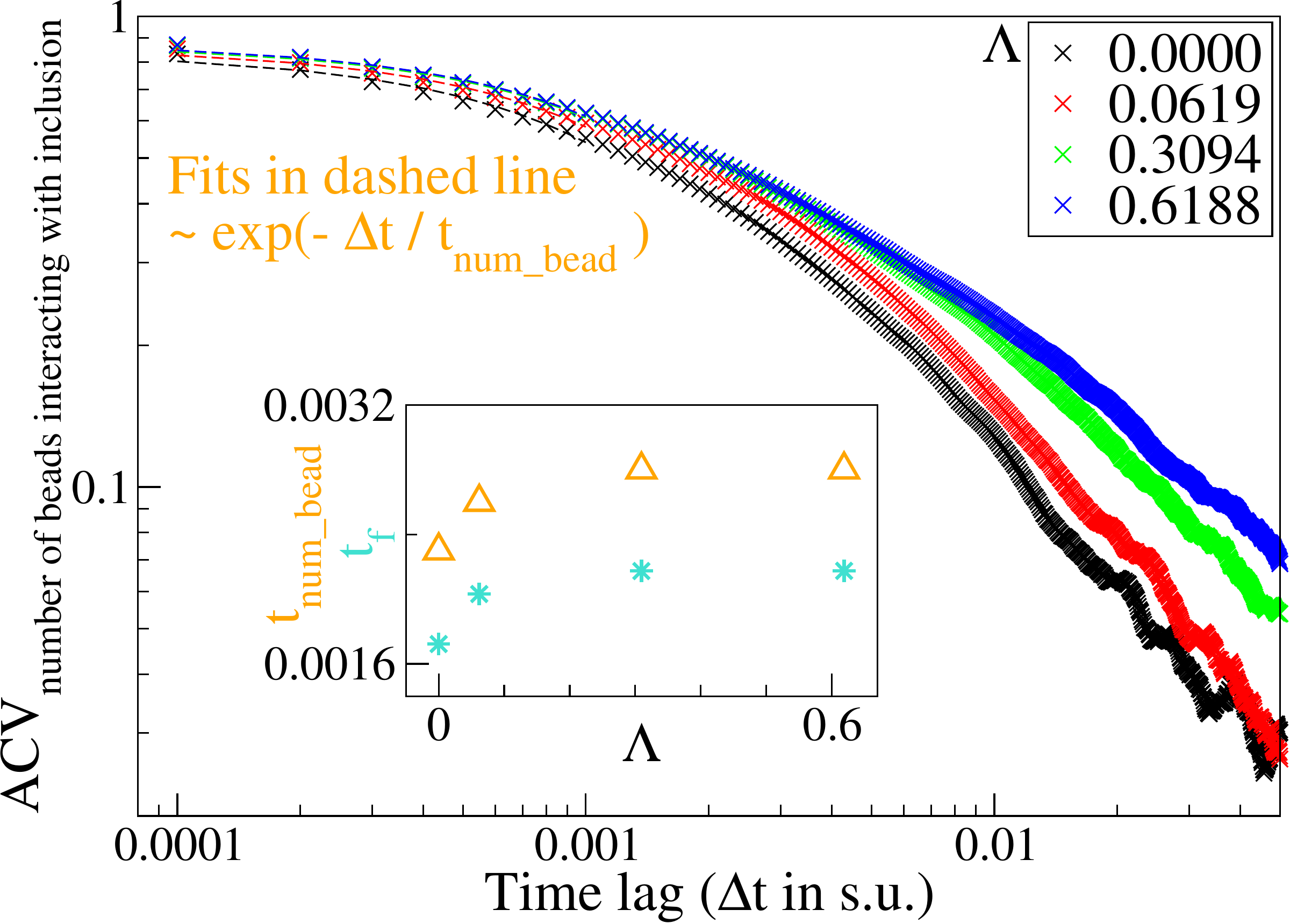} \\
\vspace{5mm}
\includegraphics[width=0.43\linewidth]{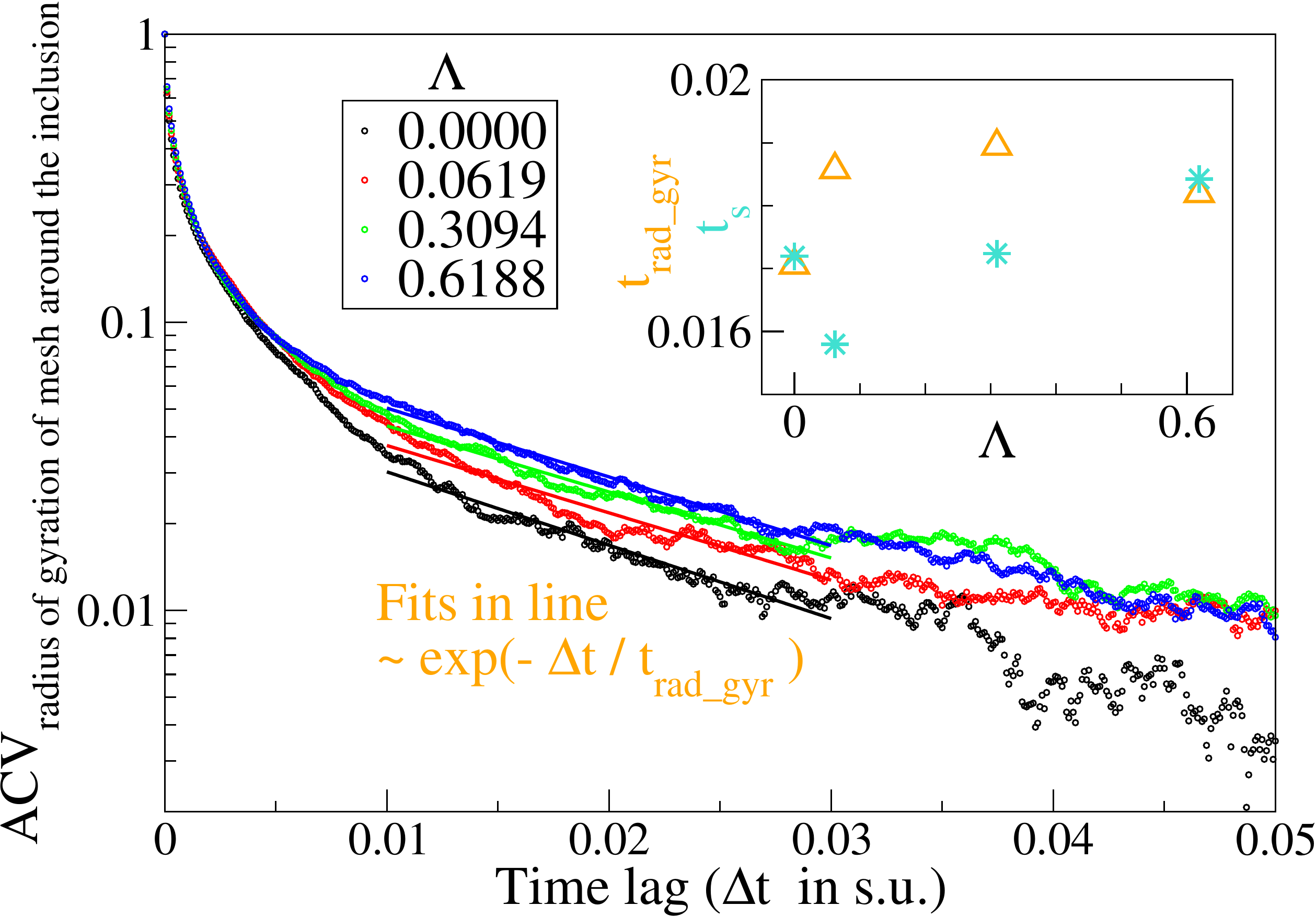}
\hspace{5mm}
\includegraphics[width=0.43\linewidth]{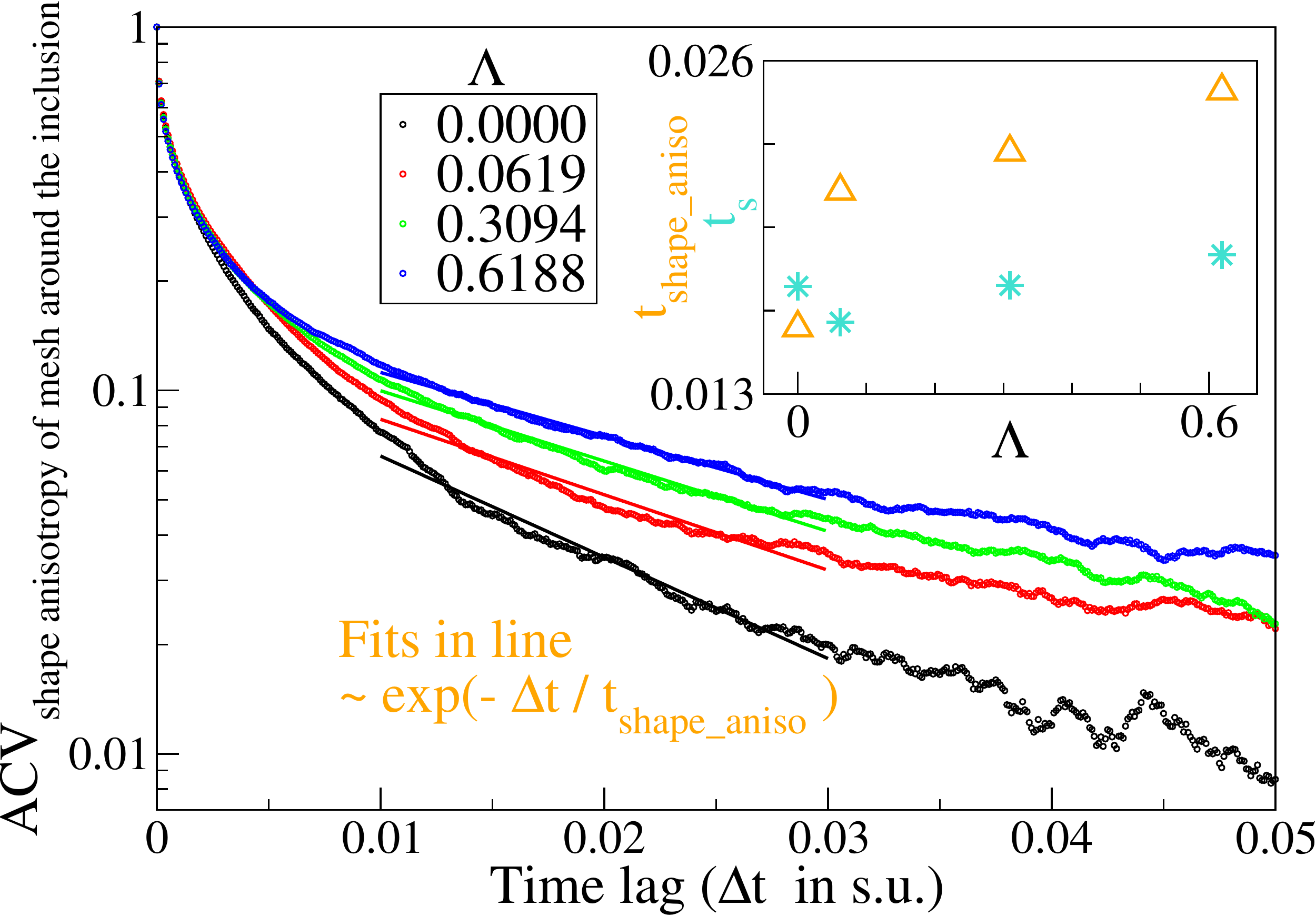}
\caption{ 
{Extraction of the fast and the slow timescales from autocovariance (ACV) of several quantities.}
(Top) ACV of the number of the polymeric beads interacting with the inclusion are fitted with exponential functions at the early lag regime. The corresponding characteristic timescales compare well with the fast timescales ($t_f(\Lambda)$). 
ACV of the radius of gyration (Bottom-left) and the shape anisotropy (Bottom-right) of polymeric mesh around the inclusion are fitted with exponential functions at the moderate lag regime. The corresponding characteristic timescales compare well with the slow timescale ($t_s(\Lambda)$). 
Data shown for $d_I = 0.40$ s.u., averaged over six realizations. 
}
\label{SFig_ACV_modes_origin}
\end{figure}

\begin{figure}[t]
\includegraphics[width=0.5\linewidth]{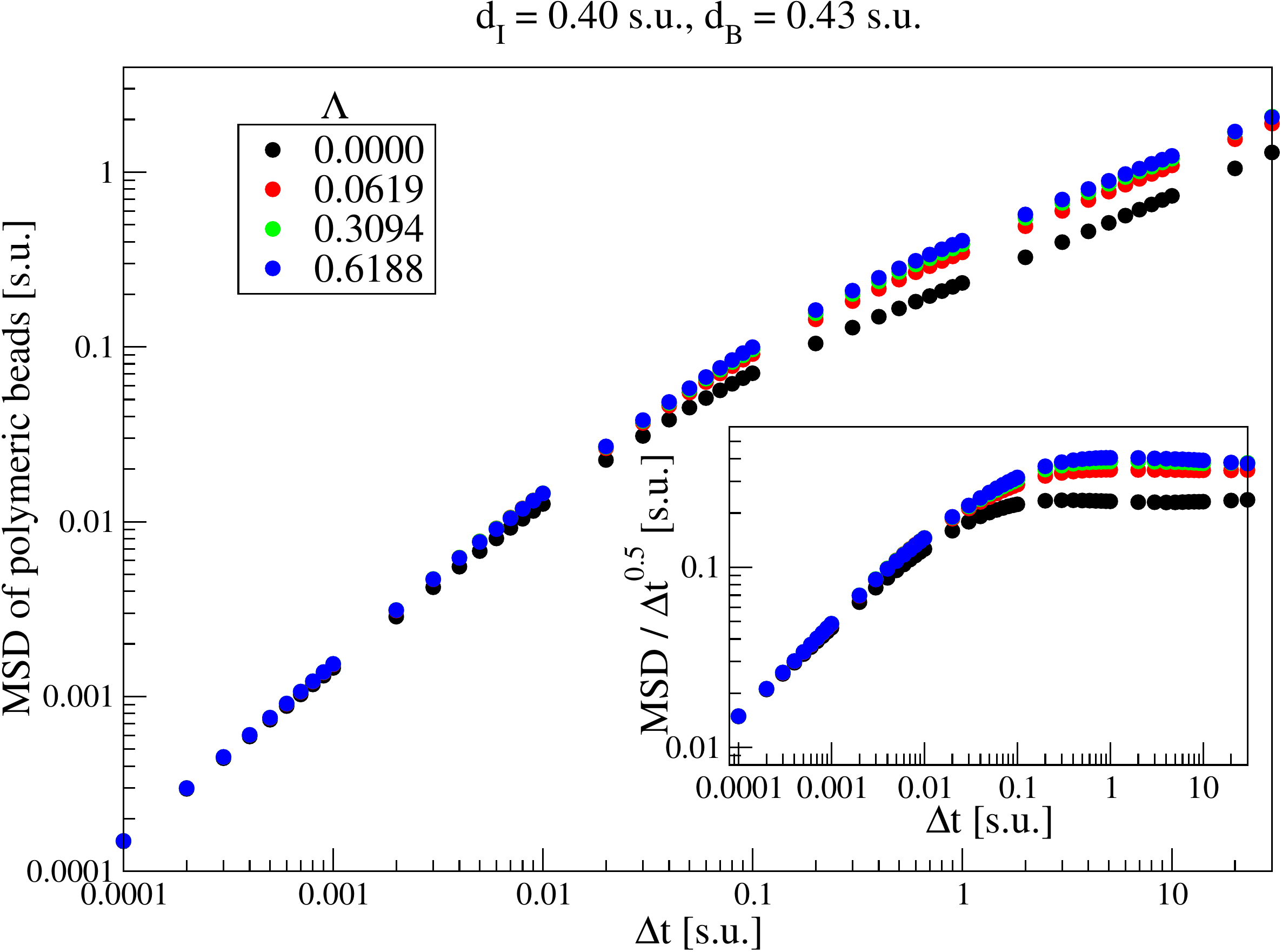}
\caption{ 
{MSD of polymeric beads.}
Individual bead diffuses normally (dynamic exponent $1$) in an early-time regime until it feels its neighbor along the polymer chain. Beyond this, the bead needs to move along with its neighbor which slows down the dynamics of the individual bead. The dynamic exponent in this regime is known to be $0.5$.
}
\label{SFig_MSD_poly}
\end{figure}

To investigate the fast and the slow modes, we tracked the beads interacting with the inclusion for all the timeframes in our simulation. 
Therefore, we could count the number of the beads interacting with the inclusion and analyze the size and the shape of the polymeric mesh around it. 
In SFIG.~\ref{SFig_ACV_modes_origin}, we show the autocovariance (ACV) of (Top) the number of beads interacting with the inclusion, (Bottom-left) the radius of gyration ({\it i.e.}, size) of the mesh, and (Bottom-right) shape anisotropy of the mesh. 
The reason behind these choices of the quantities is that any fluctuation in these quantities should induce fluctuation in the interaction potential between the inclusion and the beads surrounding it. 
We find good exponential fitting of the ACV of the number of beads interacting with the inclusion for early-time regime, giving us characteristic times for several $\Lambda$. 
Interestingly, the corresponding timescales compare well with the fast timescales obtained previously. 
Therefore, we conclude that the fast mode 
corresponds to the fluctuation in the number of polymeric beads interacting with the inclusion. 
We check the MSD of the polymeric beads and note that over this short timescale $\sim t_f$, the individual beads show normal diffusion because they are yet to realize the constraints due to their neighbors along the polymer chain (SFIG.~\ref{SFig_MSD_poly}). 
This suggests that $t_f \sim \mathcal{O}(\text{fluctuation at individual bead level})$. 
Thus, we argue that the fluctuation in the number of the beads interacting with the inclusion at such a short timescale is introduced by random inclusion or exclusion of individual bead from the the polymeric mesh around the inclusion.  

We construct gyration tensor for each timeframe using the coordinates of the beads interacting with the inclusion ({\it i.e.}, the constituents of the mesh) and, by diagonalizing that tensor, we find the radius of gyration and the shape anisotropy of the mesh. 
We show ACV of these quantities in SFIG.~\ref{SFig_ACV_modes_origin}\textendash Bottom and fit those curves with exponential functions for the moderate time regime. 
We could find considerably good fitting to those curves leading to characteristic timescales that compare well with the slow timescales ($t_s(\Lambda)$). 
Since, these quantities are related with the configuration of the mesh around the inclusion, we argue that slow mode 
corresponds to the reconfiguration aspect of the polymeric mesh.

\subsection{$\Lambda$-dependency of the diffusivities corresponding to the three dynamical modes}
\begin{figure}[t]
\includegraphics[width=0.5\linewidth]{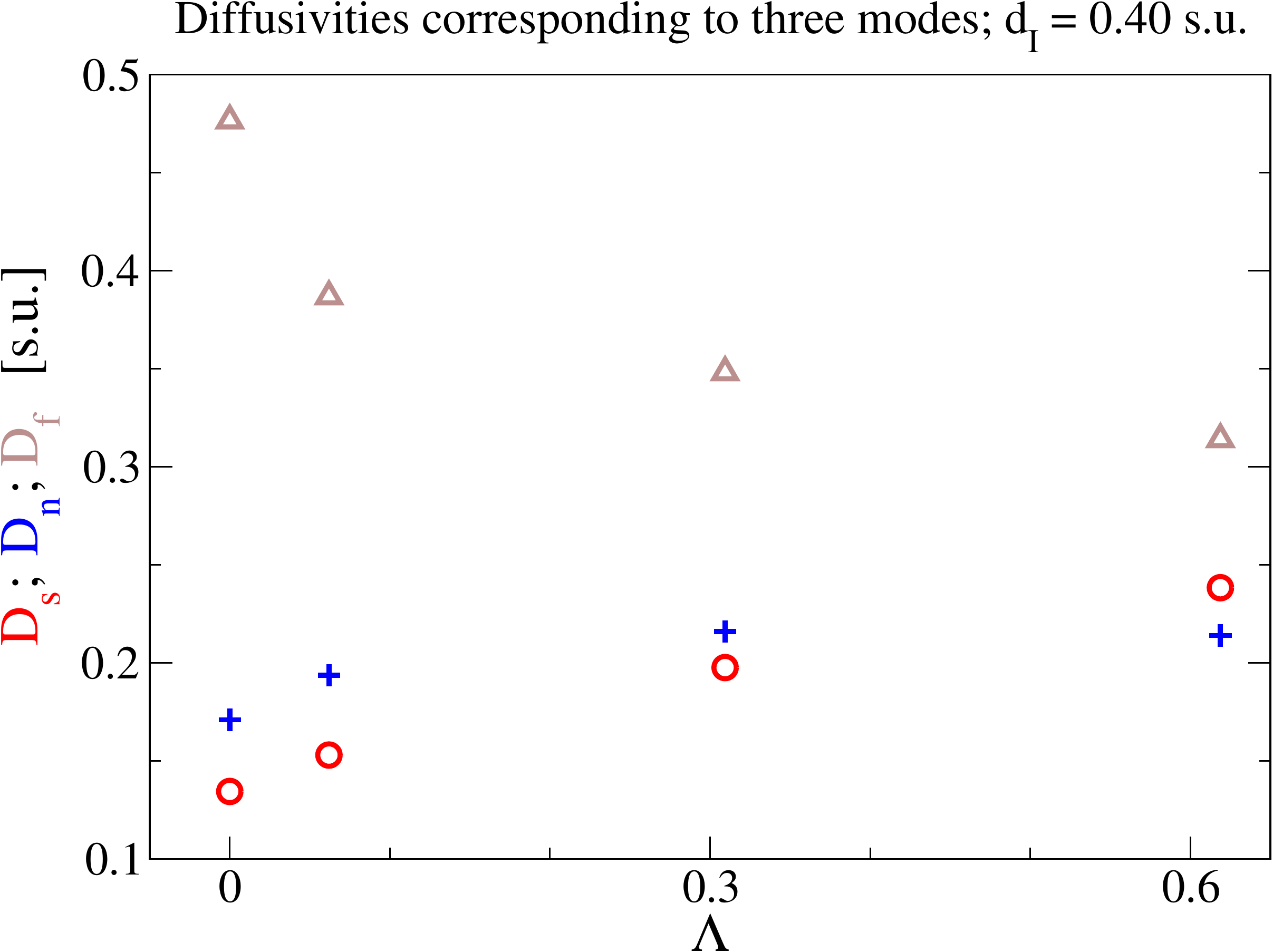}
\caption{ 
$\Lambda$-dependency of the diffusivities corresponding to the normal (blue plus symbols), fast (brown triangles), and slow modes (red circles).}
\label{SFIG_CGEM_HarmonicModel_comparison}
\end{figure}
In this section, we discuss how the three modes proposed in this article alter the inclusion dynamics with TLPA.
We show the parts of $\langle \Delta r_{I,\, EM}^2 \rangle$ derived from all the three modes in FIG.~3b\textendash main 
and the changes in the corresponding diffusivities with $\Lambda$ in 
SFIG.~\ref{SFIG_CGEM_HarmonicModel_comparison}. 
It is intuitive to understand that the fluctuation in the beads increases with $\Lambda$ (also see SFIG.~\ref{SFig_MSD_poly} to compare bead's MSD for several $\Lambda$). At short timescale $\sim t_f(\Lambda)$, this enhanced fluctuation should be randomly reverting the inclusion's motion within the mesh and thereby slowing down the inclusion dynamics.  This justifies the decrease in $D_f$ with $\Lambda$.  However, since the fast mode-part becomes irrelevant after a short timescale $t_f$, we do not see any significant $\Lambda$-dependency of this contribution to $\langle \Delta r_{I,\, EM}^2 \rangle$. On the contrary to the fast mode, we note a significant increase in $D_s$ with $\Lambda$ resulting in significant speeding up of the inclusion dynamics. This infers that TLPA-assisted remodeling of the chromatic neighborhood plays a major role in enhancing the inclusion dynamics. Also, $D_n$ increases with $\Lambda$, which is rather intuitive to understand, as the TLPA is expected to enhance fluctuation in the chromatic medium which is encapsulated within $D_n$. Note that this also enhances inclusion dynamics by a considerable extent.

\begin{figure}[b]
\includegraphics[width=0.5\linewidth]{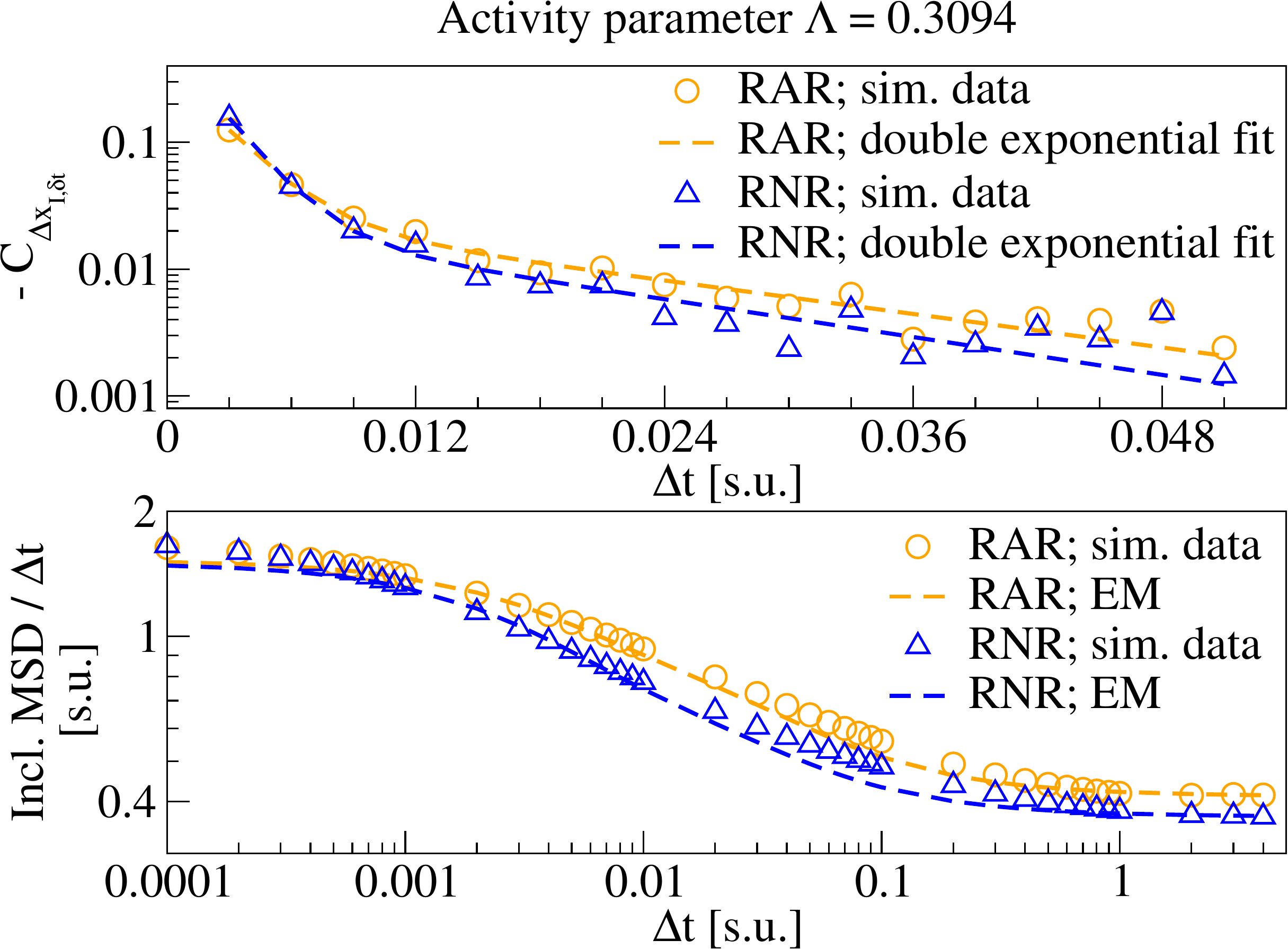}
\caption{ 
Effective model has been employed for the RAR and RNR models.} 
\label{SFIG_RAR_RNR_ACF_MSD_Lamb0_3094}
\end{figure}


\subsection{Power of the proposed effective model}
Here we illustrate the capacity of the proposed phenomenological effective model in recognizing the dynamical contributions associated with the distinct modes. 
To do so, we use the following two types of variants of our original active model:
\begin{itemize}
    \item {\bf RAR model}\textemdash 
    the steric interaction between a pair of proximal beads are subjected to the following sequence of Poisson transitions: 
    $\text{state}\left(h_{vex}>0\right) \xrightarrow{\lambda_{1}} \text{state}\left(h_{vex}<0\right) \xrightarrow{\lambda_{2}} \text{state}\left(h_{vex}>0\right)$.
    \item {\bf RNR model}\textemdash
    the steric interaction between a pair of proximal beads are subjected to the following sequence of Poisson transitions: 
    $\text{state}\left(h_{vex}>0\right) \xrightarrow{\lambda_{1}} \text{state}\left(h_{vex}=0\right) \xrightarrow{\lambda_{2}} \text{state}\left(h_{vex}>0\right)$.
\end{itemize}
The activity parameters are defined for both these models as $\Lambda = \lambda_1/\lambda_2$. 
Recall that in our original active model, the enzymatic activity is realized by the following sequence of Poisson transitions of the steric interaction between a pair of proximal beads: $\text{state}\left(h_{vex}>0\right) \xrightarrow{\lambda_{ra}} \text{state}\left(h_{vex}<0\right) \xrightarrow{\lambda_{an}} \text{state}\left(h_{vex}=0\right) \xrightarrow{\lambda_{nr}} \text{state}\left(h_{vex}>0\right)$, and the activity parameter is defined as $\Lambda = \lambda_{ra}\left( 1/\lambda_{an} + 1/\lambda_{nr} \right)$. 
For the above two variants of the original active model, we set $1/\lambda_2 = \left( 1/\lambda_{an} + 1/\lambda_{nr} \right)$ and tune $\lambda_1$ to set the corresponding activity parameter. 
Rest of the details of these variant models are kept exactly the same as in the original active model.

\begin{figure}[h]
(a) \includegraphics[width=0.5\linewidth]{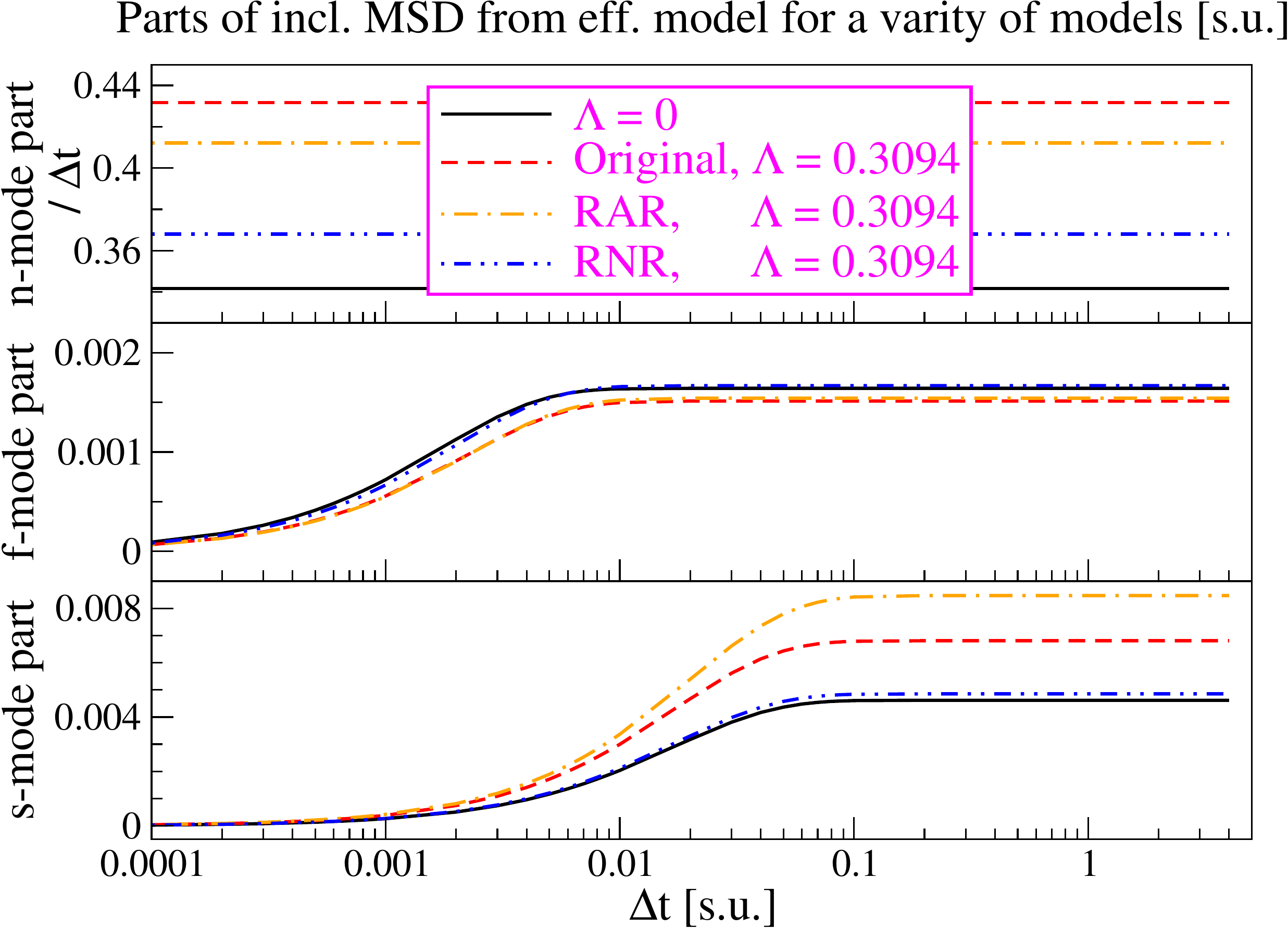} \\
(b) \includegraphics[width=0.5\linewidth]{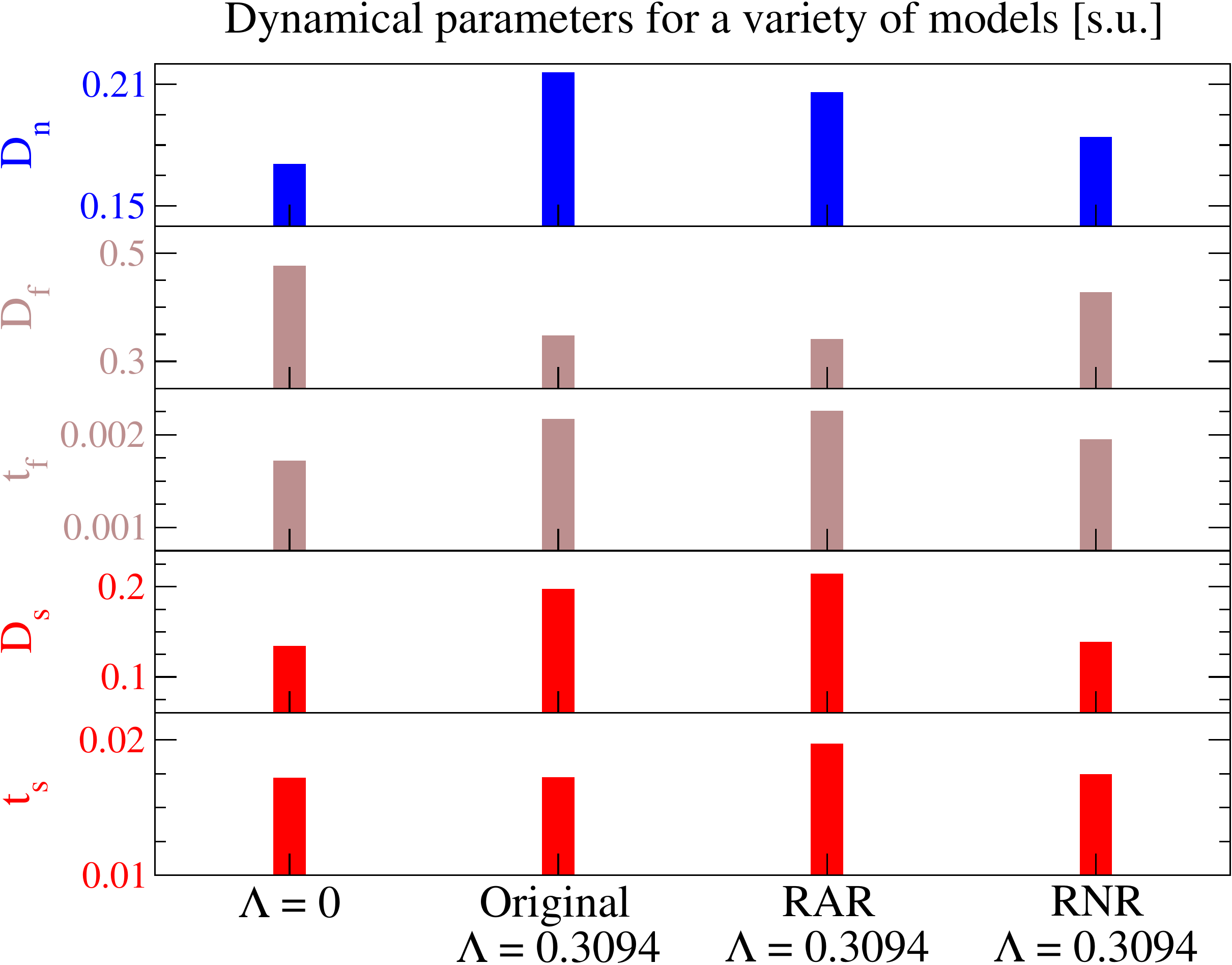}
\caption{ 
Comparison between a variety of models in terms of their dynamical modes.} 
\label{SFIG_MSD_parts_different_model}
\end{figure}

We simulate inclusion dynamics ($d_I=0.4\,\ell$) for both the RAR and RNR models with $\Lambda=0.3094$ and employ the phenomenological effective model to analyze the simulation data in the same fashion as described in the main text.
We find that our effective model successfully reproduces the simulation MSD data of the inclusion (SFIG.~\ref{SFIG_RAR_RNR_ACF_MSD_Lamb0_3094}). 
We use the corresponding dynamical parameters obtained by the effective model-scheme and determine the contributions of the distinctive dynamical modes to the total MSD of the inclusion (SFIG.~\ref{SFIG_MSD_parts_different_model}a). 
We also calculate diffusivities corresponding to the three dynamical modes and compare those with the results obtained from our original model (SFIG.~\ref{SFIG_MSD_parts_different_model}b).
Thus, we realize that our proposed effective model is significantly powerful in (i) representing the dynamics of the complex systems like those presented here in terms of a comparatively easier model and (ii) identifying the dynamical parameters associated with different physical origins (viz., dynamics within a polymeric mesh, dynamics due to mesh reconfiguration, and overall normal diffusive dynamics), which may differ from each-other across different model-scenarios. We note that a coarse-grained physical quantity like MFPT is unable to distinguish the system dynamics between the cases of the original model with $\Lambda=0$ (MFPT = $64.1364 \pm 5.2413$), RAR model (MFPT = $65.5864 \pm 4.8428$), and RNR model (MFPT = $63.6359 \pm 4.9206$). This observation further decorates the utility of our phenomenological effective model in differentiating apparently similar model-dynamics in terms of their details. 
It will be a fascinating task to analytically formulate FPT distribution using our effective model (which represents a non-Markovian process) \cite{SMVoituriezNatChem2012, SMVoituriezNature2016, SMTakahiroPRR2023}, which may explain the apparent similarity of these different model-scenarios in terms of MFPT.

\subsection{Inclusion dynamics for several inclusion sizes}
\begin{figure}[t]
\includegraphics[width=0.7\linewidth]{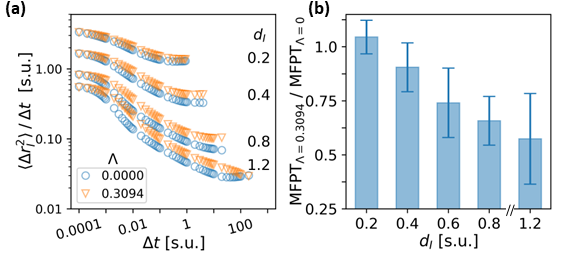}
\caption{ 
(a) MSD of the inclusion of different sizes (indicated on the right margin; typical meshsize $d_{mesh} \simeq 2\, max\left(s_{min} \vert P(s_{min})>0\right) - d_B \simeq 0.37\,\ell$ (FIG.~2c)). 
The downward shift of the curves with increasing $d_I$ is due to the corresponding mobility parameter set by Stokes' law. 
(b) Ratio of the MFPT for $\Lambda=0.3094$ to that for $\Lambda=0$ are shown for several inclusion sizes. 
The activity-dependent remodeling of the chromatic environment becomes more significant for bigger inclusion.
$n=91, \,71, \,36, \,71, \,20$ realizations are used for $d_I = 0.2, \,0.4, \,0.6, \,0.8, \,1.2 \, \ell$, respectively.
The error bars are obtained by propagating s.e.m. of the numerators and the denominators.  
}
\label{Fig4}
\end{figure}

We simulate the original APM for $d_I\in[0.2\,\ell,1.2\,\ell]$.
The MSD profiles for all the $d_I$'s investigated look similar to that we reported in the main text for $d_I=0.4\,\ell$ (SFIG.~\ref{Fig4}a). 
However, we note that the exponent of the subdiffusive regime approaches $\sim 0.5$ with increasing $d_I$ (exponents are $\sim 0.85$ and $\sim 0.5$ for $d_I = 0.2\,\ell$ and $1.2\,\ell$, respectively). 
We further note that the MSD of the polymeric beads in the corresponding time regime has an exponent $0.5$ (SFIG.~\ref{SFig_MSD_poly}). 
These two observations mutually corroborates as the constrained-dynamics (i.e., the subdiffusive behavior) of an inclusion bigger than the chromatic meshsize must be dictated by the dynamics of the neighboring beads.
Similar to what we reported in the main text, the crossover between the early-time diffusion and the intermediate subdiffusion is delayed by $\Lambda$ for all the $d_I$'s.

We also analyze the FPT statistics of the inclusion for all the $d_I$'s mentioned above. 
We note that the effect of TLPA is insignificant for the smallest $d_I$; however, it becomes significant for larger inclusions (SFIG.~\ref{Fig4}b).
For the largest $d_I$ investigated, for some $\Lambda>0$, the MFPT is reduced to almost the half its value without TLPA. 

\end{document}